\documentclass[fleqn,usenatbib]{mnras}

\usepackage{newtxtext,newtxmath}
\usepackage[T1]{fontenc}
\usepackage{ae,aecompl}

\usepackage{graphicx}
\usepackage{amsmath}
\usepackage{bm}
\usepackage{comment}

\newcommand{\beq}{\begin{equation}}
\newcommand{\eeq}{\end{equation}}
\newcommand{\beqa}{\begin{eqnarray}}
\newcommand{\eeqa}{\end{eqnarray}}

\DeclareMathOperator*{\argmax}{argmax}

\newcommand{\Msun}{\mathrm{M}_{\sun}}

\title[tSZ-WL cross-correlation with \textit{Planck} and HSC]
{Cross-correlation of the thermal Sunyaev--Zel'dovich effect and
weak gravitational lensing: \textit{Planck} and Subaru Hyper Suprime-Cam first-year data}

\author[K. Osato et al.]{
Ken Osato,$^{1,2}$\thanks{E-mail: ken.osato@iap.fr (KO)}
Masato Shirasaki,$^{3}$
Hironao Miyatake,$^{4,5,6}$
Daisuke Nagai,$^{7,8}$
\newauthor
Naoki Yoshida,$^{2,6}$
Masamune Oguri,$^{2,6,9}$ and
Ryuichi Takahashi$^{10}$
\\
$^{1}$Institut d'Astrophysique de Paris, Sorbonne Universit\'e, CNRS, UMR 7095, 75014 Paris, France\\
$^{2}$Department of Physics, Graduate School of Science, The University of Tokyo,
Bunkyo, Tokyo 113-0033, Japan\\
$^{3}$Division of Theoretical Astronomy, National Astronomical Observatory of Japan,
Mitaka, Tokyo 181-8588, Japan\\
$^{4}$Institute for Advanced Research, Nagoya University, Nagoya, Aichi 464-8601, Japan\\
$^{5}$Division of Particle and Astrophysical Science, Graduate School of Science,
Nagoya University, Nagoya, Aichi 464-8602, Japan\\
$^{6}$Kavli Institute for the Physics and Mathematics of the Universe (WPI),
The University of Tokyo Institutes for Advanced Study,\\
The University of Tokyo, Kashiwa, Chiba 277-8583, Japan\\
$^{7}$Department of Physics, Yale University, New Haven, CT 06520, USA\\
$^{8}$Yale Center for Astronomy and Astrophysics, Yale University, New Haven, CT 06520, USA\\
$^{9}$Research Center for the Early Universe, The University of Tokyo,
Bunkyo, Tokyo 113-0033, Japan\\
$^{10}$Faculty of Science and Technology, Hirosaki University, Hirosaki,
Aomori 036-8561, Japan
}

\date{Accepted XXX. Received YYY; in original form ZZZ}

\pubyear{2019}

\begin{document}
\label{firstpage}
\pagerange{\pageref{firstpage}--\pageref{lastpage}}
\maketitle

\begin{abstract}
Cross-correlation analysis of the thermal Sunyaev--Zel'dovich (tSZ) effect
and weak gravitational lensing (WL) provides a powerful probe of
cosmology and astrophysics of the intra-cluster medium.
We present the measurement of the cross-correlation of tSZ and WL
from \textit{Planck} and Subaru Hyper-Suprime Cam.
The combination enables us to study cluster astrophysics at high redshift.
We use the tSZ-WL cross-correlation and the tSZ auto-power spectrum measurements
to place a tight constraint on the hydrostatic mass bias,
which is a measure of the degree of non-thermal pressure support in galaxy clusters.
With the prior on cosmological parameters derived from
the analysis of the cosmic microwave background anisotropies by \textit{Planck}
and taking into account foreground contributions both in the tSZ auto-power spectrum
and the tSZ-WL cross-correlation, the hydrostatic mass bias is
estimated to be $26.9^{+8.9}_{-4.4} \%$ ($68\%$ C.L.),
which is consistent with recent measurements by mass calibration techniques.
\end{abstract}




\section{Introduction}
\label{sec:introduction}
The anisotropies of the cosmic microwave background (CMB) contain
rich information on the energy content and evolution of our Universe.
The so-called secondary CMB anisotropies, that are generated after last scattering,
convey further information of the large-scale structures of the Universe.
The Sunyaev--Zel'dovich (SZ)
effect \citep{Sunyaev1972,Sunyaev1980} is the most important source
of the secondary anisotropies, and it has been emerging as a powerful
observational probe into the large-scale structure.

The thermal Sunyaev--Zel'dovich (tSZ) effect is caused by
hot electrons contained in galaxy clusters,
and it has been used to study the thermodynamic properties of the intra-cluster medium (ICM).
Recently, \textit{Planck} has detected
the tSZ effect to a number of galaxy cluster with a high significance level \citep{Planck2015XXVII},
and several ground-based CMB experiments such as
Atacama Cosmology Telescope \citep[ACT,][]{Swetz2011}
and South Pole Telescope \citep[SPT,][]{Bleem2012}
measured the tSZ effect with higher angular resolution.

Several analytical prescriptions have been proposed
to model theoretically the tSZ effect and the evolution of ICM
\citep[e.g.,][]{Makino1998,Komatsu2001,Bode2009,Shaw2010,Flender2017}.
The complexities of highly non-linear ICM physics
make it challenging for such models to provide accurate theoretical predictions.
Cosmological hydrodynamical simulations are often employed to investigate the formation
and the evolution of the hot ICM
\citep[e.g.,][]{Nagai2006,Battaglia2012a,Battaglia2012b,McCarthy2014,Dolag2016}.
Unfortunately, hydrodynamical simulations are computationally expensive,
and there still remain statistical uncertainties owing to the limited simulation
volume or to the small number of samples.
In order to circumvent these problems, more observation-based approaches are proposed.
Hydrodynamics simulations and self-similar models suggest that
thermodynamic quantities such as temperature and pressure have universal profiles \citep{Nagai2007b}.
Such a profile can be expressed by a fixed functional form with
some free parameters, which are directly calibrated
through X-ray and tSZ observations \citep{Arnaud2010,PlanckIntermediateV}.
Note that the calibration can be performed for massive and nearby clusters,
but often, mostly for simplicity, a universal pressure profile model is
adopted in cosmological analysis of the tSZ effect.
One of the critical assumptions usually made is hydrostatic equilibrium (HSE);
the ICM is in dynamical equilibrium supported solely by the thermal pressure.
With the HSE assumption, we can derive the mass of galaxy clusters
from X-ray or tSZ observations in a simple manner, and can perform a variety of
cosmological analyses.

Recent cosmological simulations
show that galaxy clusters at high redshifts are not in a dynamically equilibrium state,
and are often supported by the so-called non-thermal pressure
\citep[e.g.,][]{Nelson2014b,Shi2015,Vazza2018},\footnote{The non-thermal pressure
generally includes contributions from turbulent gas motions, cosmic rays, and magnetic fields,
but the contributions of cosmic-rays and magnetic fields are estimated to be sub-dominant
from gamma-ray and radio observations, respectively.}
in addition to the thermal pressure of ICM.
Non-thermal pressure in the outer part of galaxy clusters has not yet been observed directly,
and thus the cluster mass estimates based on the HSE assumption remain uncertain or inaccurate
\citep[e.g.,][]{Nagai2007a,Lau2009,Lau2013,Suto2013,Nelson2014a,Shi2016,Biffi2016,Henson2017}.
It is important to understand the origin and the contribution of non-thermal pressure
in the next decade when large cosmological surveys are conducted;
inaccurate cluster masses may lead to biased inference of
cosmological parameters \citep[][for a recent review]{Pratt2019}.

The auto-power spectrum of the tSZ effect is widely used
as a summary statistic for cosmological analyses.
It is known that the amplitude of the tSZ power spectrum
is very sensitive to the amplitude of the matter fluctuation,
i.e., $\sigma_8$ \citep{Komatsu2002}.
At the angular-scales accessible by current observations,
most of the cosmological information from the power spectrum is in
its amplitude, and cosmological parameter inference suffers from the
well-known degeneracy
between parameters, e.g., matter density and $\sigma_8$ \citep{Bolliet2018}.
A promising way to overcome this problem is cross-correlating
the tSZ effect with another observable which traces the large-scale structure.
Since the tSZ effect traces the large-scale structure through the pressure field
in the Universe, it is expected that the cross-correlation can be detected
at high significance level.
In this work, we focus on the cross-correlation between the tSZ effect and
weak gravitational lensing (WL), which is referred to as small distortion of
images of background galaxies due to the gravitational potential generated by
large-scale structure.
The advantage of WL is that it can probe the line-of-sight integral
of density contrast directly. We do not need to introduce uncertain galaxy bias.
Various imaging surveys have successfully detected WL with high signal-to-noise ratio
in sufficiently wide areas for cosmological studies \citep{Kuijken2015,Zuntz2018}.

The cross-correlation of tSZ and WL has already been detected \citep{VanWaerbeke2014,Hojjati2017}
and utilised to probe into the diffuse gas distribution \citep{Ma2015}.
\citet{Osato2018} use the cross-correlation to constrain both
cosmological parameters and the fractional contribution from non-thermal pressure.
However, the capabilities of the measurements so far are limited
due to low surface number density of source galaxies detected by shallow observations.
In the present paper, we use the WL measurements
by Subaru Hyper-Suprime Cam \citep{Miyazaki2018,Aihara2018a,Mandelbaum2018a}.
The high image quality and the large aperture enable us to
detect faint galaxies at high redshift.
We can extract the information of
the large-scale structure and ICM physics at high redshift
that is not accessible by other WL surveys.

The rest of the paper is organized as follows.
In Section~\ref{sec:formulation}, we describe the method to predict
the auto-power spectrum of tSZ and the cross-correlation
of tSZ and WL.
Then, we describe WL survey by Hyper Suprime-Cam in Section~\ref{sec:WL_HSC},
and tSZ observation by \textit{Planck} in Section~\ref{sec:tSZ_Planck}.
In Section~\ref{sec:mock_observations}, details of mock simulations used for estimation
of covariance matrix are described.
We present the measurement of the cross-correlation in Section~\ref{sec:measurement}
and the cosmological analyses in Section~\ref{sec:analyses}.
In Section~\ref{sec:discussions}, discussions on the results are presented and
the concluding remarks are made in Section~\ref{sec:conclusions}.

\section{Formulation}
\label{sec:formulation}
In this Section, we present our analytic model
to generate theoretical templates of the tSZ auto-power spectrum
and the tSZ-WL cross-correlation.

\subsection{The thermal Sunyaev--Zel'dovich effect}
Here, we briefly review the basics of the tSZ effect.
The temperature variation due to the tSZ effect
is proportional to the line-of-sight integral of
the electron pressure with respect to the physical distance $l$,
\beq
\frac{\Delta T}{T_0} = g_\nu (x) y = g_\nu (x) \frac{\sigma_\mathrm{T}}{m_\mathrm{e} c^2}
\int  \!\! P_\mathrm{e} \, dl,
\eeq
where $T_0 = 2.725 \, \mathrm{K}$ is the temperature of CMB photons,
$y$ is the Compton-$y$ parameter,
$\sigma_\mathrm{T}$ is the Thomson scattering cross-section,
$m_\mathrm{e}$ is the electron mass,
$c$ is the speed of light, and $P_\mathrm{e}$ is the electron pressure.
The function $g_\nu (x)$ determines the dependence of the frequency $\nu$:
\beq
g_\nu (x) = x \frac{e^x - 1}{e^x + 1} - 4, \ x = \frac{h\nu}{k_\mathrm{B} T_0} ,
\eeq
where $h$ is the Planck constant and $k_\mathrm{B}$ is the Boltzmann constant.
For very hot or relativistic electrons,
relativistic corrections may become important \citep{Itoh1998,Nozawa1998,Chluba2012}
but we do not consider the typically small corrections in the following cosmological
analysis.

\subsection{Power spectrum of the tSZ effect}
\label{sec:auto_halomodel}
Since the Compton-$y$ is the integration of the product of density and temperature,
the main contribution of tSZ comes from hot gas of clusters and
the contribution from diffuse gas in the cluster outskirts and in filaments is subdominant.
In order to model the power spectrum of Compton-$y$, we employ the so-called halo model,
where all matter is associated with halos \citep{Makino1993,Komatsu1999,Shaw2010}.

The power spectrum can be decomposed into contributions from a single halo
and clustered two halos, which are referred to as one-halo and two-halo terms,
respectively.
Then, the auto-power spectrum $C^{yy} (\ell)$ of the Compton-$y$ parameter is given as
\beq
\label{eq:yy_hm}
C^{yy} (\ell) = C^{yy}_\mathrm{1h}  (\ell)+ C^{yy}_\mathrm{2h}  (\ell) .
\eeq
The one-halo term is expressed as
\beq
\label{eq:yy_1h}
C^{yy}_\mathrm{1h} (\ell) = \int \! \! dz \frac{d^2V}{dz d\Omega}
\int \! \! dM \frac{dn (M,z)}{dM}
|\tilde{y}_\ell (M,z)|^2 ,
\eeq
where $d^2 V / dz d\Omega = D_A^2(z) (c/H(z))$ is the comoving volume per redshift
and solid angle, $D_A (z)$ is the comoving angular diameter distance, and
$dn (M, z) / dM$ is the halo mass function.
The 2D Fourier transform of the Compton-$y$ parameter $\tilde{y}_\ell (M, z)$ is given as
\beq
\tilde{y}_\ell (M, z) = \frac{4 \pi r_s}{\ell_s^2}
\left( \frac{\sigma_\mathrm{T}}{m_e c^2} \right)
\int \!\! dx \, x^2 P_\mathrm{e} (x; M, z) \frac{\sin (\ell x / \ell_s)}{\ell x / \ell_s},
\eeq
where $r_s$ is the arbitrary scale radius, $\ell_s = (1+z)^{-1} D_A (z) /r_s$, and
$P_\mathrm{e} (x; M,z)$ is the electron pressure profile
with respect to the scaled radius $x \equiv r/r_s$.
We define the halo radius $R_\Delta$ as the radius within which
the mean density is equal to $\Delta$ times the critical density $\rho_\mathrm{cr} (z)$.
Then, the enclosed mass $M_\Delta$ is given as
\beq
\label{eq:enclosed_mass}
M_\Delta = \frac{4\pi}{3}\Delta \rho_\mathrm{cr} (z) R_\Delta^3 .
\eeq
We adopt the virial halo mass $M_\mathrm{vir}$, which overdensity $\Delta_\mathrm{vir}$
is computed from spherical collapse model \citep{Bryan1998},
\beq
\label{eq:Delta_vir}
\Delta_\mathrm{vir} = 18 \pi^2 + 82 (\Omega_\mathrm{m} (z) - 1)
- 39 (\Omega_\mathrm{m} (z) - 1)^2 ,
\eeq
where $\Omega_\mathrm{m} (z)$ is the matter density normalized by the critical density
at redshift $z$:
\beq
\Omega_\mathrm{m} (z) = \Omega_\mathrm{m} (1+z)^3 E^{-2}(z) ,
\eeq
and $E(z)$ is the expansion factor:
\beq
E(z) = \frac{H(z)}{H_0} = [\Omega_\mathrm{m} (1+z)^3 + 1 - \Omega_\mathrm{m}]^{\frac{1}{2}} .
\eeq
Here, $\Omega_\mathrm{m}$ is the present value of matter density
normalized by the critical density and
we assume the flat $\Lambda$ cold dark matter (CDM) Universe.
There are alternative halo mass definitions.
One is the critical overdensity masses, $M_{200}$ and $M_{500}$,
which mean density is equal to $200$ and $500$ times
the critical density $\rho_\mathrm{cr} (z)$, respectively.
Furthermore, we also adopt the mean overdensity mass $M_\mathrm{200m}$,
which mean density is equal to $200$ times the background matter density $\rho_\mathrm{m} (z)$.
Note that all masses can be converted interchangeably
once the mass and the concentration parameter (see Section~\ref{sec:cross_halomodel})
are specified for one mass definition.
Throughout halo model calculations,
we use $M_\mathrm{vir}$ as the definition of the halo mass $M$.

Next, the two-halo term is given by
\beqa
\label{eq:yy_2h}
C^{yy}_\mathrm{2h} (\ell) &=& \int \!\! dz \frac{d^2V}{dz d\Omega}
P_\mathrm{m} \left( k=\frac{\ell+1/2}{D_A (z)}, z \right) \nonumber \\
&& \times \left[ \int \!\! dM \frac{dn (M, z)}{dM}
\tilde{y}_\ell (M, z) b_\mathrm{h} (M, z) \right]^2,
\eeqa
where $b_\mathrm{h} (M, z)$ is the halo bias and $P_\mathrm{m} (k,z)$
is the 3D linear matter power spectrum.
Finally, we can compute the power spectrum of Compton-$y$,
if the halo mass function, halo bias, cosmological parameters, and pressure profile
are specified.
We adopt the fitting formulas of halo mass function of \citet{Bocquet2016} and
halo bias of \citet{Tinker2010}, where the mean overdensity mass $M_\mathrm{200m}$
is adopted as the halo mass definition.
We use the linear Boltzmann code \texttt{CAMB} \citep{Lewis2000}
to compute the linear matter power spectrum.

In computing the Fourier transform of the Compton-$y$ parameter $\tilde{y}_\ell (M, z)$,
we adopt the universal pressure profile of \citet{Nagai2007b}. The explicit form is given by
\beqa
\frac{P_\mathrm{e}(r)}{P_{500}} &=& p(x)
\left[ \frac{M^\mathrm{HSE}_\mathrm{500}}{3\times 10^{14} h_{70}^{-1} \Msun } \right]^{0.12}, \\
p(x) &\equiv& \frac{P_0}{(c_{500}x)^\gamma
[1+(c_\mathrm{500}x)^\alpha]^{(\beta-\gamma)/\alpha}} , \\
P_{500} &=& 1.65 \times 10^{-3} E(z)^{8/3} \nonumber \\
&& \times \left[ \frac{M^\mathrm{HSE}_\mathrm{500}}{3\times 10^{14} h_{70}^{-1} \Msun} \right]^{2/3}
h_{70}^2 \,\mathrm{keV}\,\mathrm{cm}^{-3},
\eeqa
where $x = r/R^\mathrm{HSE}_{500}$ and $h_{70} = h/0.7$.
This formula contains several free parameters, which are directly fitted to data from
X-ray or SZ observations of the cluster pressure profile.
With the measurements of SZ selected clusters \citep{PlanckIntermediateV},
the parameters are calibrated as
\beq
(P_0, c_{500}, \gamma, \alpha, \beta) = (6.41, 1.81, 0.31, 1.33, 4.13) .
\eeq
Note that the cluster mass is estimated with HSE assumption
and thus is likely underestimated compared with the true mass.
In order to relate the HSE mass with true mass,
we introduce the hydrostatic bias parameter $b_\mathrm{HSE}$
and scale the mass and radius as
$M^\mathrm{HSE}_\mathrm{500} = M_\mathrm{500} (1-b_\mathrm{HSE})$
and $R^\mathrm{HSE}_{500} = R_{500} (1-b_\mathrm{HSE})^{1/3}$.
Typically, $b_\mathrm{HSE} = 0.15 \text{--} 0.40$ is estimated from
WL mass calibration measurements and also from hydrodynamical simulations
(see Sections~\ref{sec:comp_obs} and \ref{sec:comp_sim} for more details and references therein).

\subsection{Cross-correlations of tSZ and WL}
\label{sec:cross_halomodel}
Similarly to the auto-power spectrum,
the cross-power spectrum of Compton-$y$ and convergence field
is also computed based on halo model:
\beqa
C^{y\kappa} (\ell) &=& C^{y\kappa}_\mathrm{1h} (\ell) +
C^{y\kappa}_\mathrm{2h} (\ell) ,
\label{eq:yk_hm} \\
C^{y\kappa}_\mathrm{1h} (\ell) &=& \int \!\! dz \frac{d^2V}{dz d\Omega}
\int \!\! dM \frac{dn (M,z)}{dM}
\tilde{y}_\ell (M,z) \tilde{\kappa}_\ell (M,z) ,
\label{eq:yk_1h} \\
C^{y\kappa}_\mathrm{2h} (\ell) &=& \int \!\! dz \frac{d^2V}{dz d\Omega}
P_\mathrm{m} \left( k=\frac{\ell+1/2}{D_A(z)}, z \right) \nonumber \\
&& \times \int \!\! dM \frac{dn (M, z)}{dM}
\tilde{y}_\ell (M, z) b_\mathrm{h} (M, z) \nonumber \\
&& \times \int \!\! dM \frac{dn (M, z)}{dM}
\tilde{\kappa}_\ell (M, z) b_\mathrm{h} (M, z) .
\label{eq:yk_2h}
\eeqa
Compared with the formulas of the auto-power spectrum
(Eqs.~\ref{eq:yy_hm}, \ref{eq:yy_1h}, and \ref{eq:yy_2h}),
one of the Fourier transform of Compton-$y$ $\tilde{y}_\ell (M, z)$ is replaced with
the Fourier transform of the convergence from a single halo $\tilde{\kappa}_\ell (M, z)$.
Before moving onto the explicit expression of $\tilde{\kappa}_\ell (M, z)$,
we review the basics of the halo density profile, which is critical to model the lensing signal.
It is assumed that the density profile
has a spherical profile, so-called Navarro--Frenk--White (NFW) profile
\citep{Navarro1996,Navarro1997}:
\beq
\rho (r) = \frac{\rho_s}{(r/r_s) (1+r/r_s)^2} ,
\eeq
where $\rho_s$ is the scale density and $r_s$ is the scale radius.
Since this density profile scales as $\propto r^{-3}$ at large radii,
the enclosed mass does not converge.
Thus, we truncate the profile at the virial radius $R_\mathrm{vir}$
determined through Eqs.~\eqref{eq:enclosed_mass} and \eqref{eq:Delta_vir}.
Then, the virial halo mass is given as
\beq
M_\mathrm{vir} = \int_0^{R_\mathrm{vir}} \!\! \rho (r) 4\pi r^2 dr =
4\pi \rho_s r_s^3 m_\mathrm{NFW} (c),
\eeq
where
\beq
m_\mathrm{NFW} (c) = \int_0^c \frac{r}{(1+r)^2} \, dr = \ln (1+c) - \frac{c}{1+c} ,
\eeq
and $c = R_\mathrm{vir}/r_s$ is the concentration parameter.
In order to determine the profile, we need to specify the scale radius $r_s$
and the scale density $\rho_s$.
It is known that the scale radius is correlated with the halo mass,
and the fitting formula of the concentration parameter
as a function of virial mass and redshift is used.
Throughout this paper, we adopt the fitting formula
in \citet{Klypin2016} calibrated with $N$-body simulations.\footnote{In \citet{Klypin2016},
the fitting formula is given as the function of the virial mass and
the free parameters are tabulated for redshifts of simulation outputs.
In order to obtain the concentration parameter at arbitrary redshift,
we linearly interpolate these parameters.
For details, refer to Table~A3 of \citet{Klypin2016}.}
As a result, for a given virial mass, the density profile is uniquely determined.
Then, the Fourier transform of the scaled density profile $u(r) = \rho(r)/M$ is given as
\beqa
\tilde{u}_M (k) &\equiv& \int d^3 r \, u (r) e^{- i \bm{k} \cdot \bm{r}} \\
&=& \frac{1}{m_\mathrm{NFW} (c)}
\left[ \sin x \{ \mathrm{Si} [x(1+c)] - \mathrm{Si} (x) \} \right. \nonumber \\
&& + \cos x \{ \mathrm{Ci} [x(1+c)] - \mathrm{Ci} (x) \} \nonumber \\
&& \left. - \frac{\sin (xc)}{x(1+c)} \right] ,
\eeqa
where $x \equiv (1+z) k r_s$, $\mathrm{Si} (x)$ and $\mathrm{Ci} (x)$
are sine and cosine integrals, respectively \citep{Scoccimarro2001}.
The Fourier transform of the convergence profile $\kappa (\theta)$
from a single halo with mass $M$ is given by
\beq
\tilde{\kappa}_\ell (M, z) = \int 2 \pi \theta \kappa(\theta) J_0 (\ell \theta) \, d\theta
= \frac{M\tilde{u}_M(k = \ell/D_A(z), z)}{D_A^2(z) \Sigma_\mathrm{cr}(z)} ,
\eeq
where $J_0 (x)$ is the zeroth order Bessel function.
The critical surface mass density $\Sigma_\mathrm{cr} (z)$ is given as
\beq
\Sigma_\mathrm{cr}^{-1} (z) = \frac{4 \pi G}{c^2}
\chi(z) (1+z)^{-1} \left[ 1 - \chi(z)
\left\langle \frac{1}{\chi (z_s)} \right\rangle \right] ,
\eeq
where $\chi (z)$ is the comoving distance and
\beq
\left\langle \frac{1}{\chi (z_s)} \right\rangle = \left[ \int dz_s \frac{d p}{dz_s} \frac{1}{\chi(z_s)} \right]
\left[ \int dz_s \frac{d p}{dz_s} \right]^{-1} .
\eeq
The stacked probability distribution function (PDF) of source galaxy redshifts $p(z)$
needs to be given by the actual catalog and only the stacked PDF depends on the observational data
in this prescription.
By summing up PDFs of all source galaxies, dilution effect is mitigated
though it is not completely removed \citep[see also][]{Medezinski2018}.
In the subsequent analysis, we adopt \texttt{Ephor AB} with reweights
derived from COSMOS 30 band observations
as our fiducial choice (see Section~\ref{sec:photo-z}).

\subsection{Differential contribution with respect to halo mass and redshift}
\label{sec:differential_signal}
\begin{figure*}
\includegraphics[width=\textwidth]{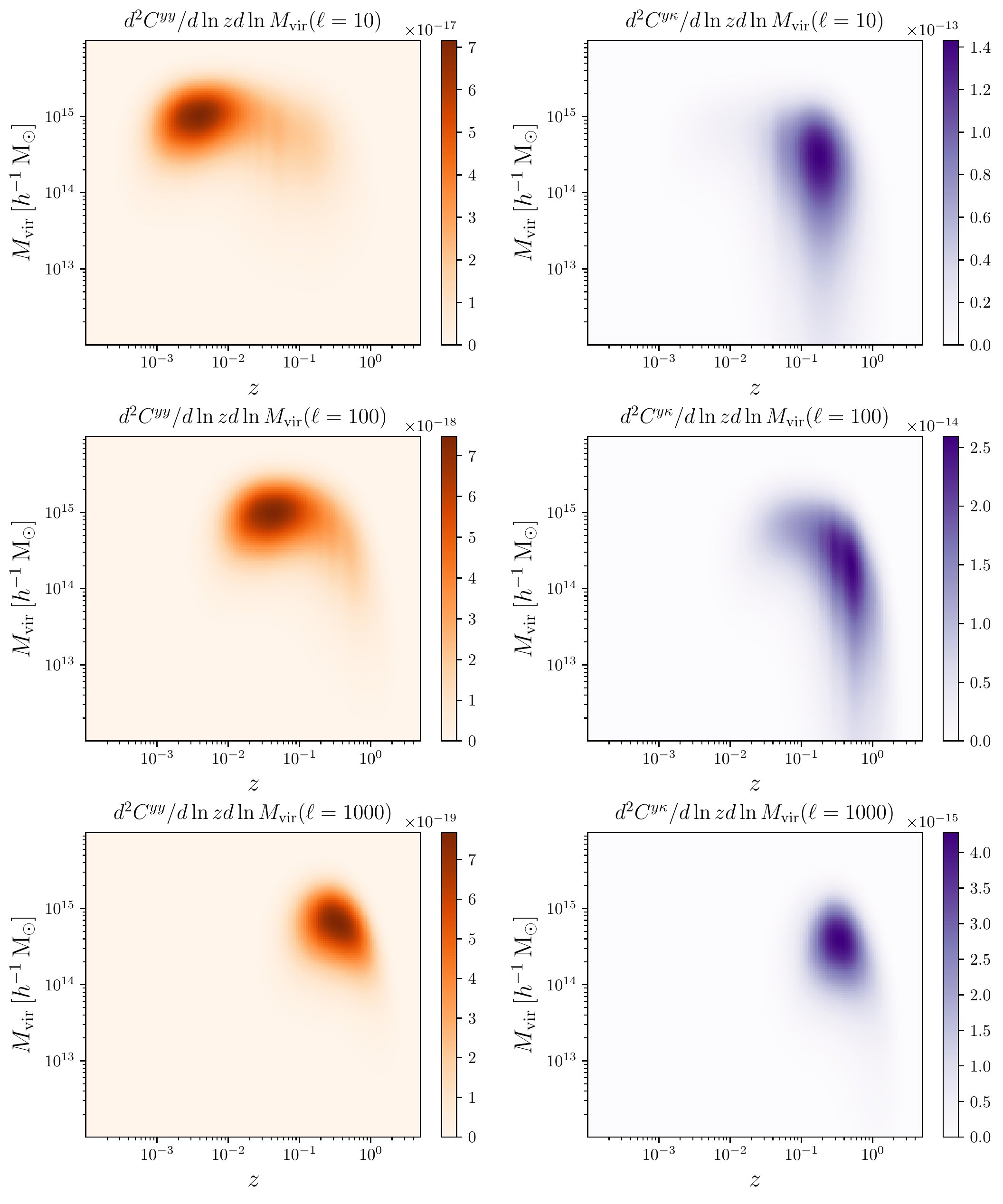}
\caption{Differential contributions with respect to halo mass
and redshift for tSZ auto-power spectrum (left panels)
and tSZ-WL cross-power spectrum (right panels).
In each row, the contributions for multipoles $\ell = 10, 100, 1000$
are shown. Note that the plotted differential contributions are with respect to
$\ln z$ and $\ln M_\mathrm{vir}$, instead of $z$ and $M_\mathrm{vir}$.}
\label{fig:dCl}
\end{figure*}

In the halo model, the tSZ auto-power spectrum and tSZ-WL cross-power spectrum
have been sourced by halos with various ranges of mass and redshift.
To investigate the mass and redshift ranges
of which halos contribute to the spectra the most,
we compute the differential contribution for both of spectra.
From Eqs.~\eqref{eq:yy_hm}, \eqref{eq:yy_1h}, and \eqref{eq:yy_2h},
the differential contribution for tSZ auto-power spectrum with respect to
the halo mass and redshift is given by
\beqa
\frac{d^2 C^{yy} (\ell)}{dzdM} &=&
\frac{d^2 C^{yy}_\mathrm{1h} (\ell)}{dzdM} +
\frac{d^2 C^{yy}_\mathrm{2h} (\ell)}{dzdM}, \\
\frac{d^2 C^{yy}_\mathrm{1h} (\ell)}{dzdM} &=&
\frac{d^2V}{dz d\Omega} \frac{dn (M,z)}{dM}
|\tilde{y}_\ell (M,z)|^2 , \\
\frac{d^2 C^{yy}_\mathrm{2h} (\ell)}{dzdM} &=&
\frac{d^2V}{dz d\Omega} P_\mathrm{m} \left( k=\frac{\ell+1/2}{D_A (z)}, z \right)
\nonumber \\
&& \times 2 \left[ \frac{dn (M, z)}{dM}
\tilde{y}_\ell (M, z) b_\mathrm{h} (M, z) \right.
\nonumber \\
&& \left. \int \!\! dM \frac{dn (M, z)}{dM}
\tilde{y}_\ell (M, z) b_\mathrm{h} (M, z) \right] .
\eeqa
Similarly for the tSZ-WL cross-power spectrum,
from Eqs.~\eqref{eq:yk_hm}, \eqref{eq:yk_1h}, and \eqref{eq:yk_2h},
the differential contribution of the cross-power spectrum is given by
\beqa
\frac{d^2 C^{y\kappa} (\ell)}{dzdM} &=&
\frac{d^2 C^{y\kappa}_\mathrm{1h} (\ell)}{dzdM} +
\frac{d^2 C^{y\kappa}_\mathrm{2h} (\ell)}{dzdM}, \\
\frac{d^2 C^{y\kappa}_\mathrm{1h} (\ell)}{dzdM} &=&
\frac{d^2V}{dz d\Omega} \frac{dn (M,z)}{dM}
\tilde{y}_\ell (M,z) \tilde{\kappa}_\ell (M,z), \\
\frac{d^2 C^{y\kappa}_\mathrm{2h} (\ell)}{dzdM} &=&
\frac{d^2V}{dz d\Omega} P_\mathrm{m} \left( k=\frac{\ell+1/2}{D_A (z)}, z \right)
\nonumber \\
&& \times \left[ \frac{dn (M, z)}{dM}
\tilde{y}_\ell (M, z) b_\mathrm{h} (M, z)
\right.
\nonumber \\
&& \times \int \!\! dM \frac{dn (M, z)}{dM}
\tilde{\kappa}_\ell (M, z) b_\mathrm{h} (M, z)
\nonumber \\
&& + \frac{dn (M, z)}{dM}
\tilde{\kappa}_\ell (M, z) b_\mathrm{h} (M, z)
\nonumber \\
&& \times \left. \int \!\! dM \frac{dn (M, z)}{dM}
\tilde{y}_\ell (M, z) b_\mathrm{h} (M, z) \right] .
\eeqa
In Figure~\ref{fig:dCl}, the differential contributions
for three multipoles $\ell = 10, 100, 1000$ are shown.
Though both of spectra are sensitive to the similar range of halo mass and redshift
for $\ell = 1000$, for other multipoles, less massive and higher redshift halos
contribute to the tSZ-WL cross-power spectrum.
This result clearly shows the tSZ auto-power spectrum and tSZ-WL cross-power spectrum
contain the information from different classes of halos.

\section{Weak lensing analysis with HSC S16A}
\label{sec:WL_HSC}

\subsection{Shape catalog}
Subaru Hyper Suprime-Cam (HSC) is the imaging camera \citep{Miyazaki2015,Miyazaki2018}
mounted on the prime focus of the Subaru telescope.
HSC has the excellent wide field of view of $1.5 \, \mathrm{deg}$ diameter,
which corresponds to $1.77 \, \mathrm{deg}^2$.
The HSC survey \citep{Aihara2018a,Aihara2018b} is composed of three layers
according to science targets and depth:
Wide, Deep, and UltraDeep. The WL analysis makes use of Wide layer data,
which will cover $1400 \, \mathrm{deg}^2$ over six years for five broad-bands, $grizy$.

The first-year HSC shape catalog, labelled as S16A, is based on observation data
taken from 2014 March to 2016 April for about 90 nights.
We employ cuts to the whole galaxy sample to create the shape catalog.
Such cuts include a \texttt{cmodel} magnitude cut $i < 24.5$
\citep[for definitions of \texttt{cmodel} magnitude in the HSC survey, see][]{Bosch2018},
in contrast to the magnitude limit of HSC ($i \sim 26.4$).
In addition, galaxies which point spread function (PSF) is failed to be estimated
are removed and the regions affected by bright stars are masked.
Thus, the shape catalog is constructed in a conservative manner.
For each galaxy which passes all the criteria, we estimate the ellipticity $\bm{e}$
with the re-Gaussianization PSF correction method \citep{Hirata2003}
for $i$-band co-added images:
\beq
\bm{e} = (e_1, e_2) = \frac{1-(b/a)^2}{1+(b/a)^2} (\cos 2\phi, \sin 2\phi) ,
\eeq
where $b/a$ is the minor-to-major axis ratio of galaxy images and
$\phi$ is the polar argument of the major axis.
The resultant shape catalog is defined for $136.9\,\mathrm{deg}^2$
and the survey region is split into six patches:
GAMA15H, WIDE12H, GAMA09H, VVDS, XMM, and HECTOMAP.

Here we describe additional quantities in the shape catalog used for our analysis.
The full production process of the shape catalog and associated image simulations
are outlined in \citet{Mandelbaum2018a} and \citet{Mandelbaum2018b}, respectively.
The S16A shape catalog has the per-component rms galaxy shape of
a galaxy population $e_\mathrm{rms}$ for each source galaxy which is calibrated
against the image simulations, and the weight $w_i$ which is defined as
the inverse-variance weight with $e_\mathrm{rms}$ and the measurement noise caused by photon noise.
The S16A galaxy shapes should be calibrated using the multiplicative bias factor $m$
and additive bias $c_\alpha$, where $m$ is shared among the ellipticity components and
$c_{\alpha}$ is defined for an ellipticity component $\alpha$.
These calibration factors are derived from the image simulations.

\subsection{Reconstruction of the convergence field}
In order to derive the weak lensing convergence field,
we employ Kaiser--Squires inversion (hereafter, KS inversion) method \citep{Kaiser1993}.
We follow the analysis of \citet{Oguri2018},
where the same shape catalog is used.
First, we estimate the shear filed from the shape of source galaxies
\beq
\hat{\gamma}_\alpha (\bm{\theta}) =
\frac{\sum_i w_i (\gamma_\alpha (\bm{\theta}_i) - c_{\alpha , i})
W_\mathrm{G}(|\bm{\theta} - \bm{\theta}_i|)}{\sum_i w_i (1+m_i) W_\mathrm{G}(|\bm{\theta} - \bm{\theta}_i|)} ,
\eeq
where $w_i$ is the lens weight, $\gamma_\alpha (\bm{\theta}_i)$ is
the local shear estimated as $\gamma_\alpha (\bm{\theta}_i) = e_{\alpha , i} / 2 \mathcal{R}$
with the galaxy shape ellipticity $e_{\alpha , i}$ and the shear responsivity $\mathcal{R}$.
Hereafter, the subscript $i$ runs over all source galaxies in the patch.
The shear responsivity $\mathcal{R}$ is evaluated as
\beq
\mathcal{R} = 1-\frac{\sum_i w_i e_{\mathrm{rms}, i}^2}{\sum_i w_i} ,
\eeq
where $e_{\mathrm{rms}, i}$ is the intrinsic shape dispersion.
The shear responsivity is evaluated for each patch.
We apply smoothing with the Gaussian kernel $W_\mathrm{G} (\theta)$:
\beq
\label{eq:Gaussian_window}
W_\mathrm{G}(\theta) = \frac{1}{\pi \theta_s^2} \exp \left( -\frac{\theta^2}{\theta_s^2} \right) ,
\eeq
where the smoothing scale $\theta_s$ is adopted as $\theta_s^\kappa = 2 \, \mathrm{arcmin}$.
This choice of the smoothing kernel ensures maps with high signal-to-noise ratio \citep{Oguri2018}.
The additive and multiplicative biases, $c_{\alpha , i}$ and $m_i$,
are calibrated with image simulations in \citet{Mandelbaum2018b}.
Then, we convert the shear field to the convergence field as
\beq
\label{eq:hat_kappa}
\hat{\kappa} (\bm{\theta}) = \frac{1}{\pi} \int d^2\theta'
\frac{\gamma_\mathrm{t} (\bm{\theta}|\bm{\theta}')}{|\bm{\theta}-\bm{\theta}'|^2},
\eeq
where $\gamma_\mathrm{t} (\bm{\theta}|\bm{\theta}')$ is the tangential shear
at the position $\bm{\theta}$ with respect to $\bm{\theta}'$.

In the practical analysis, we adopt the flat-sky approximation.
First, we compute the pixelized shear field
on the regular grid with the pixel size of $0.5 \, \mathrm{arcmin}$
for each patch of the survey regions,
where the boundaries are determined by the positions of the source galaxies.
Though the window function $W_\mathrm{G} (\bm{\theta})$ is always non-zero,
we truncate it so that the function is forced to be zero
outside the square with a side length $8 \theta_s^\kappa$ centered on each galaxy position.
Then, we apply fast Fourier transform (FFT) to the shear field
and from Eq.~\eqref{eq:hat_kappa}, the Fourier transform of convergence field is given as
\beq
\hat{\kappa} (\bm{\ell}) = \frac{(\ell_1 + i \ell_2)^2}{\ell^2}
\hat{\gamma} (\bm{\ell}).
\eeq
Then, the convergence field is obtained by inverse FFT.
Although the convergence field should be a real function,
the resultant field may have imaginary components.
The real and imaginary parts are called as $E$- and $B$-mode convergence, respectively.
In the subsequent analysis, we use only $E$-mode convergence,
which we simply refer to as the convergence field,
and the $B$-mode will be used for null tests (see Section~\ref{sec:null_tests}).

\begin{figure*}
\includegraphics[width=\textwidth]{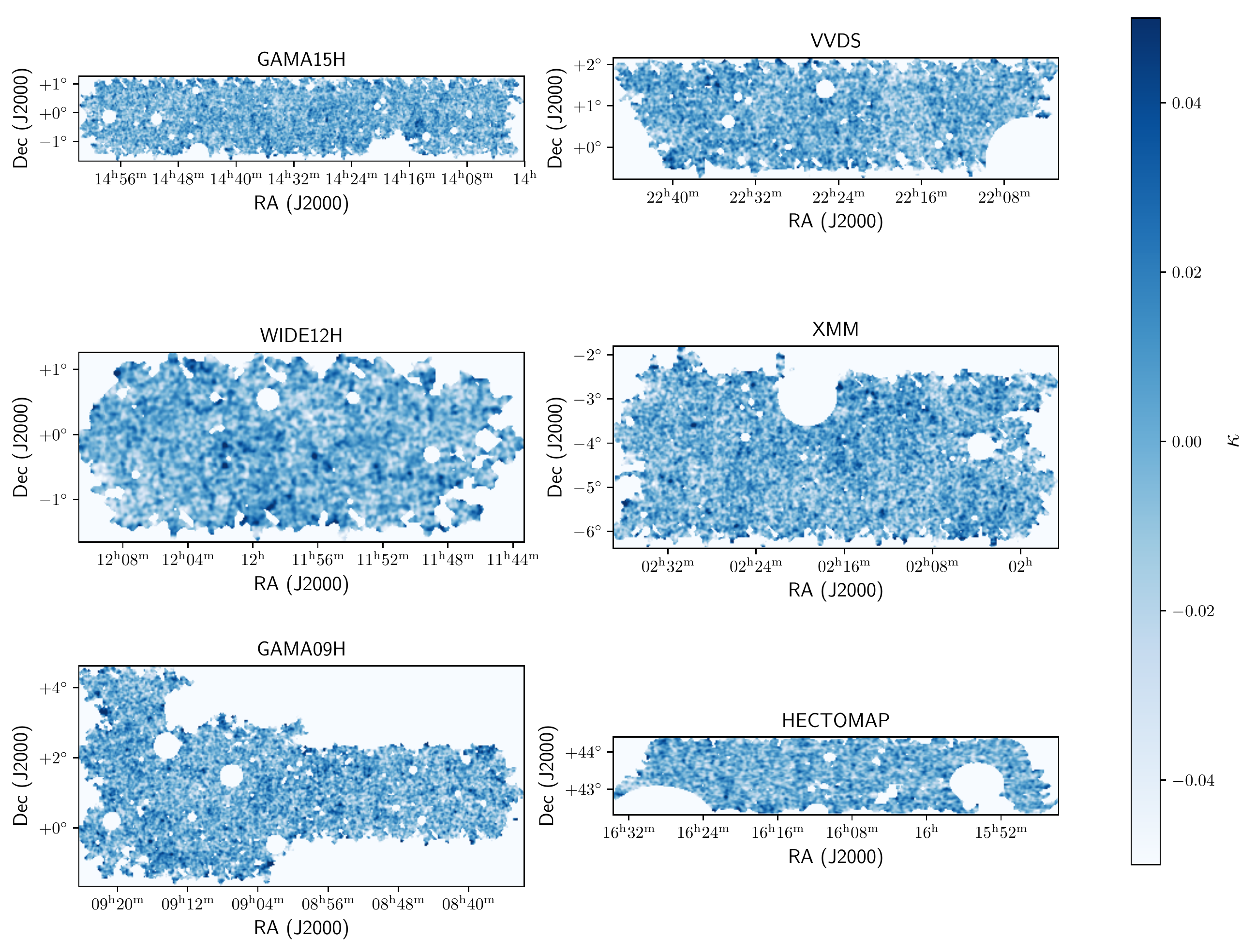}
\caption{The reconstructed convergence fields with HSC S16A shape catalogs
for six different survey patches:
GAMA15H, WIDE12H, GAMA09H, VVDS, XMM, and HECTOMAP.}
\label{fig:convergence_HSC}
\end{figure*}

We also compute smoothed number density field\footnote{Because the number density
of source galaxies is used only for determining the mask in the convergence map,
the difference between the smoothed number densities with and without lens weights
has negligible effects on the results.}:
\beq
n (\bm{\theta}) = \sum_i W_\mathrm{G} (|\bm{\theta} - \bm{\theta}_i|) .
\eeq
In order to remove the effects due to boundary and low density pixels,
we mask the pixel where the smoothed number density is less than
the half of the mean number density.
The total area after masking is $161.66 \, \mathrm{deg}^2$.
We summarize properties for each patch in Table~\ref{tab:survey_properties}
and show the reconstructed convergence field in Figure~\ref{fig:convergence_HSC}.
Note that the area of survey regions defined in shape catalogs is $136.9 \, \mathrm{deg}^2$
but the convergence map covers $161.66 \, \mathrm{deg}^2$ because of the non-local nature of reconstruction.

\begin{table*}
\caption{Number of source galaxies, area after masking,
and mean smoothed number density for six survey patches and all survey regions.}
\label{tab:survey_properties}
\begin{tabular}{cccc}
  \hline
  Field & Number of galaxies & Area ($\mathrm{deg}^2$) & Mean smoothed number density ($\mathrm{deg}^{-2}$) \\
  \hline
  GAMA15H & $2794258$ & $34.54$ & $17.90$ \\
  WIDE12H & $1219607$ & $14.31$ & $18.05$ \\
  GAMA09H & $3005495$ & $41.01$ & $15.91$ \\
  VVDS & $1843091$ & $22.25$ & $17.22$ \\
  XMM & $2596006$ & $32.47$ & $17.29$ \\
  HECTOMAP & $1157693$ & $17.09$ & $12.75$ \\
  \hline
  All fields & $12616150$ & $161.66$ & --- \\
  \hline
\end{tabular}
\end{table*}

\subsection{Source Redshift Distributions}
\label{sec:photo-z}
\begin{figure}
\includegraphics[width=\columnwidth]{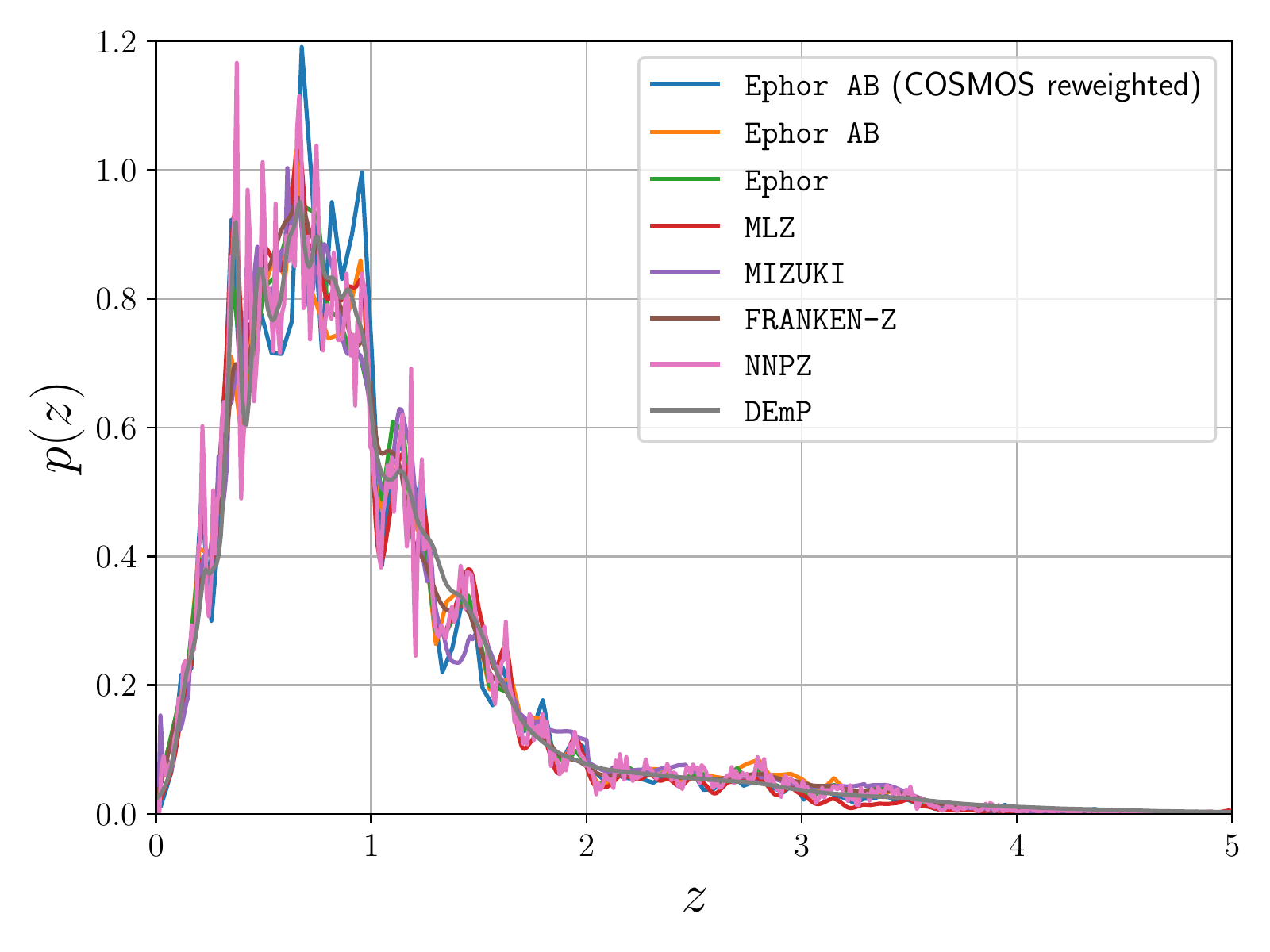}
\caption{The stacked PDFs of source galaxy redshifts for different algorithms:
\texttt{Ephor AB} with COSMOS reweight, \texttt{Ephor AB},
\texttt{Ephor}, \texttt{MLZ}, \texttt{MIZUKI}, \texttt{FRANKEN-Z}, \texttt{NNPZ},
and \texttt{DEmP}.
Our fiducial choice is \texttt{Ephor AB} with COSMOS reweight.}
\label{fig:pz}
\end{figure}

To calculate the WL convergence, we need
the stacked PDF of the source galaxy redshifts (see Section~\ref{sec:cross_halomodel}).
We sum up PDFs of all galaxies in the S16A shape catalogs.
There are a number of algorithms to estimate the photometric redshifts of
source galaxies \citep[for details, see][]{Tanaka2018}.
Figure~\ref{fig:pz} shows the stacked PDFs of source galaxy redshifts
derived using eight different algorithms.
Since HSC can detect faint galaxies,
the resultant PDF has a tail at high redshifts.
As a fiducial model, we employ the stacked PDF estimated with \texttt{Ephor AB} code
by reweighting the PDF obtained from the COSMOS 30-band observation catalog \citep{Ilbert2009,Laigle2016}
so that the distributions of magnitudes for five bands of HSC should match with
that of galaxies used in the analysis \citep[for details, see Section~5.2 of][]{Hikage2019}.
We confirm that the different algorithms give consistent results within a few per-cent
for calculations of cross-correlations (see Appendix~\ref{sec:pz_algos}).

\subsection{Blinding}
In the current situation where many cosmological results, e.g., constraints on cosmological parameters,
are available, there is a risk that if the analysis becomes consistent with other results,
the further analysis will not be carried out and
the possible systematics will not be investigated any longer.
This degrades the credibility and quality of the analysis, and is called as confirmation bias.
In order to avoid the confirmation bias and derive robust results,
we follow a blinding scheme in our analysis.
In practice, we adopt two-tiers blinding of the multiplicative bias, i.e.,
\beq
\bm{m}^\text{cat}_i = \bm{m}^\text{true} + d\bm{m}^{(1)}_i + d\bm{m}^{(2)}_i \  (i = 0, 1, 2) ,
\eeq
where $\bm{m}^\text{cat}_i$ is the array of multiplicative bias stored in the shape catalog.
The first term $d\bm{m}^{(1)}_i$ is added to avoid the case where the analysis lead accidentally finds
true catalog when one of multiple projects is unblinded.
This term is decrypted every time the catalog is used but not referenced directly.
The second term $d\bm{m}^{(2)}_i$ is the factor disclosed by a blinder after all necessary analysis
and unblinding procedure are performed.
One of $d\bm{m}^{(2)}_i \  (i = 0, 1, 2)$ is exactly zero, which corresponds to the true catalog,
and the index of the true catalog is notified to the analysis lead
from the blinder because only the blinder can decrypt $d\bm{m}^{(2)}_i$.
Once the true catalog is disclosed, all of the results are fixed
and further analysis and modification of the analysis pipeline are prohibited.
The details of the blinding scheme are found in Section~3.2 of \citet{Hikage2019}.

\section{The thermal Sunyaev--Zel'dovich effect measured by \textit{Planck}}
\label{sec:tSZ_Planck}
\begin{figure*}
\includegraphics[width=\textwidth]{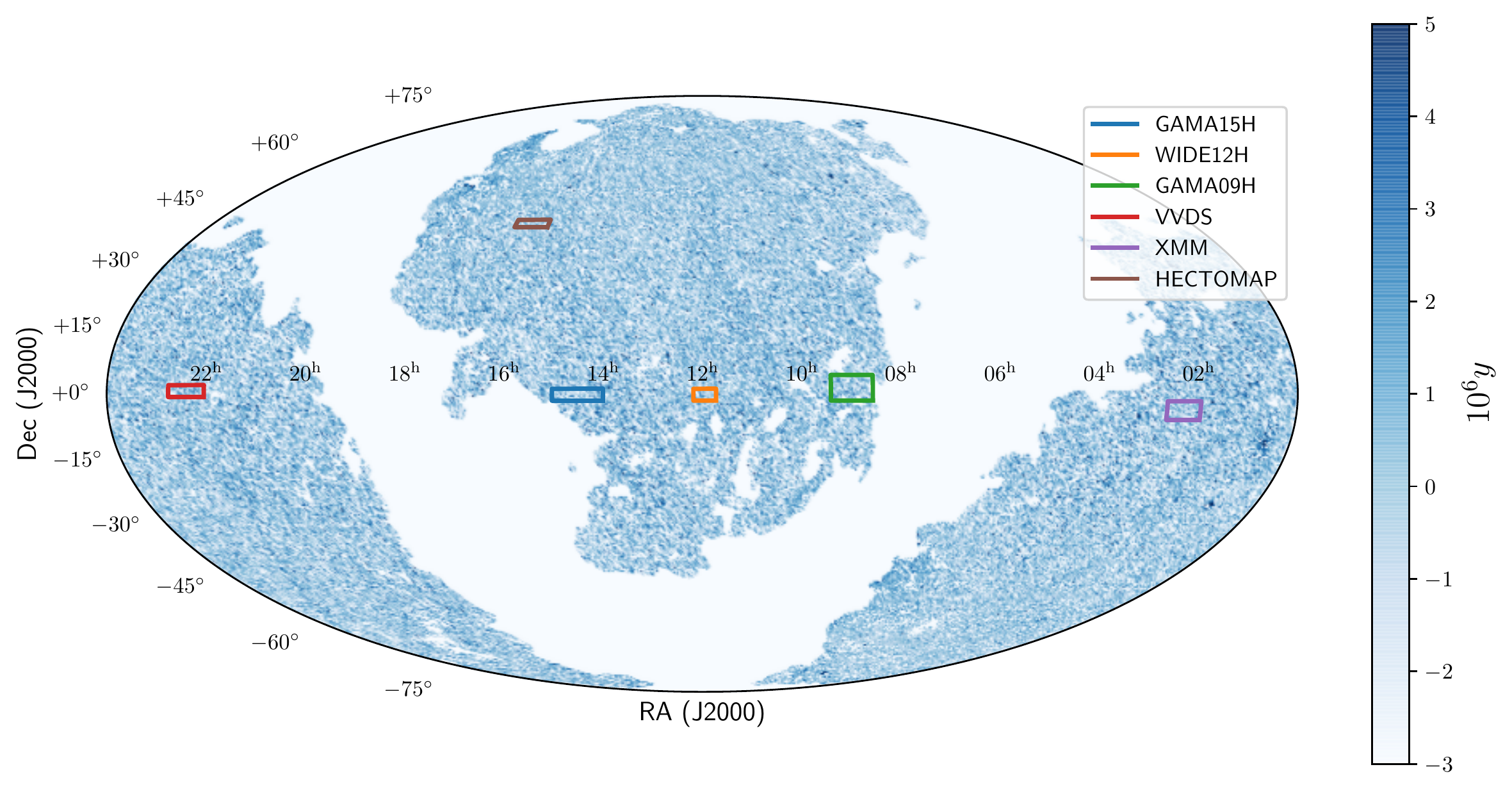}
\caption{The Compton-$y$ map measured by \textit{Planck}
based on the \texttt{MILCA} algorithm.
The six boxes indicate the HSC S16A survey patches.
The 40\% galactic mask and the point source mask have been applied.}
\label{fig:Planck_ymap}
\end{figure*}

Here, we briefly review the construction process of Compton-$y$ maps
from \textit{Planck} measurements. All of details are found in \citet{Planck2015XXII}.
The Compton-$y$ map of \textit{Planck} is constructed from $30$ to $857 \, \mathrm{GHz}$
channel maps of the \textit{Planck} full mission data
with component separation algorithm.
The map is pixelized in \texttt{Healpix} \citep{Gorski2005} format
with $N_\mathrm{side} = 2048$.
The beam properties are different between the maps observed by different bands,
but we assume circularly symmetric Gaussian beam
with the full-width half-mean (FWHM) beam size $\theta_\mathrm{FWHM} = 10.0 \, \mathrm{arcmin}$
for the Compton-$y$ map, which corresponds to the Gaussian window scale
$\theta_s^y = \theta_\mathrm{FWHM}/(2 \sqrt{\log 2}) = 6.0 \, \mathrm{arcmin}$
(see Eq.~\ref{eq:Gaussian_window}).
The \textit{Planck} team provides maps with two different component separation algorithm:
\texttt{MILCA} \citep[Modified Internal Linear Combination Algorithm,][]{Hurier2013} and
\texttt{NILC} \citep[Needlet Independent Linear Combination,][]{Remazeilles2011},
both of which basically try to find the linear combination of several components
so that the variance of the reconstructed map is minimized.
Hereafter, we use the map constructed with \texttt{MILCA} as the fiducial map
because it has lower noise contribution at large scales.
When measuring the auto-power spectrum of Compton-$y$,
the effect due to contamination originating from systematics must be minimized.
To this end, the full mission data are separated by half,
and the cross-power spectrum between the separated first and second maps
is used as a baseline power spectrum.
We make use of the full mission data for the measurement of
the cross-correlations because such systematics do not correlate with
the lensing convergence field.
In addition, we mask galactic planes and point sources,
where strong radio emission component separation becomes unreliable.
We employ the 40\% galactic mask and point source mask
provided by \textit{Planck} collaboration.
We show the \texttt{MILCA} Compton-$y$ map together with HSC S16A survey patches
in Figure~\ref{fig:Planck_ymap}.

\section{Mock observations}
\label{sec:mock_observations}
In this Section, we present details of mock observations of tSZ and WL.
We measure the tSZ auto-power spectrum and tSZ-WL cross-correlations
from mock observations, and the results are used to estimate
the covariance matrix.
We also use these mock catalogues for null tests (Section~\ref{sec:null_tests})
and evaluate the significance of
our cross-correlation measurement (Section~\ref{sec:significance}).

\subsection{All-sky mock Compton-$y$ maps}
We generate mock tSZ maps from all-sky halo catalogs of \citet{Takahashi2017}.
In the simulations, cosmological parameters are adopted
from \textit{WMAP} 9-year results \citep{Hinshaw2013}:
the CDM density parameter $\Omega_\mathrm{cdm} = 0.233$,
the baryon density parameter $\Omega_\mathrm{b} = 0.046$,
the matter density parameter $\Omega_\mathrm{m} = \Omega_\mathrm{cdm} + \Omega_\mathrm{b} = 0.279$,
the cosmological constant density $\Omega_\Lambda = 0.721$,
the scaled Hubble constant $h = 0.7$,
the amplitude of the matter power spectrum $\sigma_8 = 0.82$,
and the spectral index of the scalar perturbation $n_\mathrm{s} = 0.97$.
From the halo catalog, we construct the all-sky tSZ map
by pasting the universal pressure profile with $b_\mathrm{HSE} = 0.2$
onto each halo in the catalog.
First, we create 108 mock tSZ maps from the halo catalog.
Next, we smooth the tSZ map with the circular Gaussian window function
with the Gaussian window scale $\theta^y_s = 6.0 \, \mathrm{arcmin}$.
Finally, we apply the $40\%$ galactic and radio point source masks.
The sky coverage fraction after masking is $f_\mathrm{sky} = 0.512$.
We do not add instrumental noise to the mock maps
because the amplitude of the noise is uncertain and the foreground noise is dominant.

\begin{figure}
\includegraphics[width=\columnwidth]{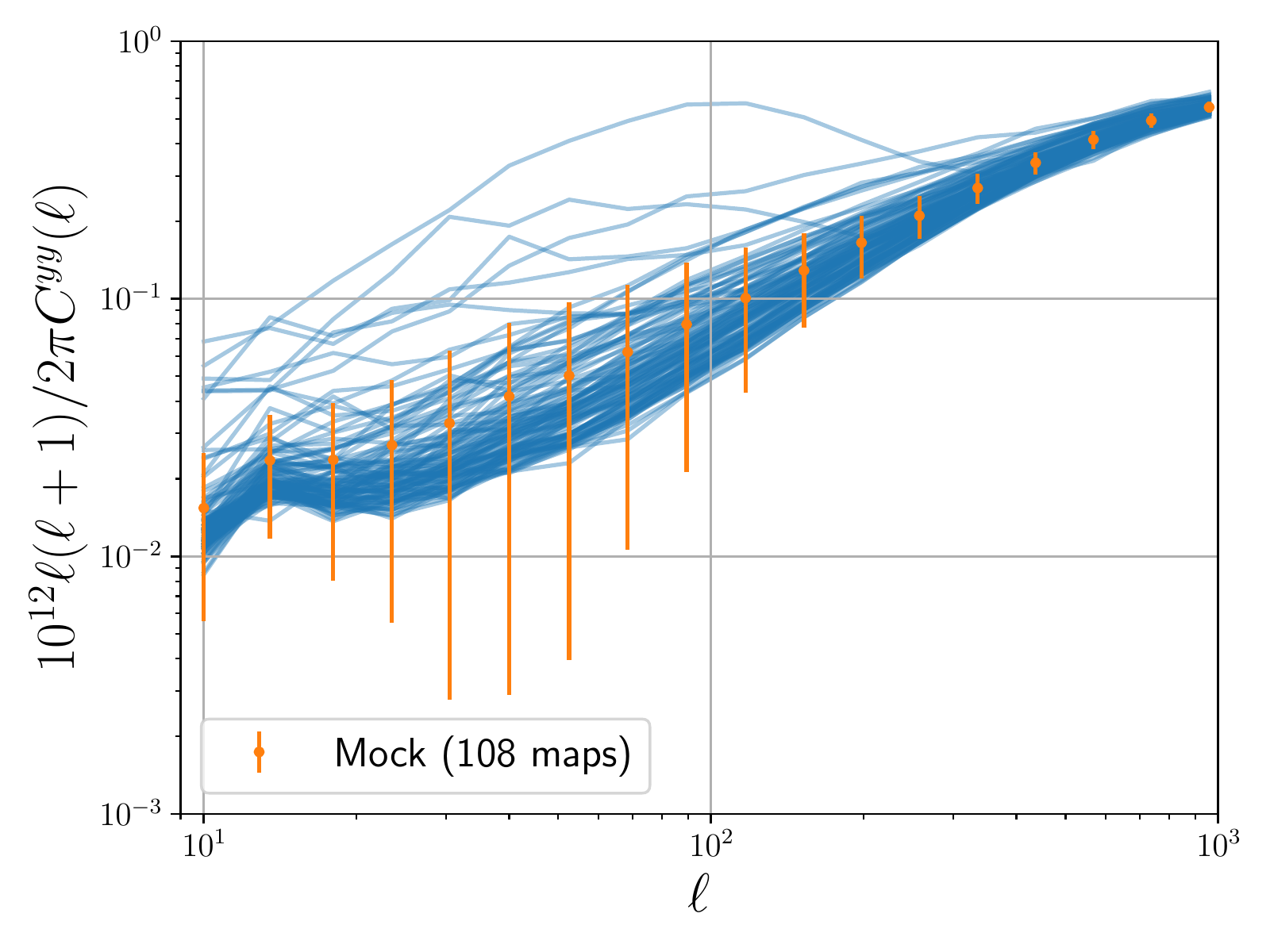}
\caption{The power spectra measured from mock Compton-$y$ maps.
Each blue solid line corresponds to one measurement with a mock map.
The orange points with error bars show the mean and standard deviation
among 108 mock maps.
Note that the standard deviation is computed for the raw power spectrum,
not logarithm of it. Thus, the lower limit of the error looks smaller
than the distribution of power spectra.
Another cause is that the distribution of power spectra at large scales
are not well approximated as Gaussian distribution.}
\label{fig:cl_mock}
\end{figure}

In order to remove the artificial mode coupling induced by masking,
we deconvolve the mask spectrum from the pseudo-spectrum
using \texttt{MASTER} algorithm \citep{Hivon2002}.
The relation between the pseudo-power spectrum $\tilde{C}_\ell$,
which is measured directly from mock maps,
and the true power spectrum $C_\ell$ can be given as
\beq
\label{eq:pseudospec}
\tilde{C}_\ell = \sum_{\ell'} M_{\ell \ell'} B^2_{\ell'} C_{\ell'},
\eeq
where $M_{\ell \ell'}$ is the mode-coupling matrix,
and $B_\ell$ is the window function describing smoothing effects of the beam
and finite pixelization of \texttt{Healpix}.
The mode-coupling matrix is given as
\beq
M_{\ell_1 \ell_2} = \frac{2 \ell_2 + 1}{4 \pi} \sum_{\ell_3} (2 \ell_3 + 1)
W_{\ell_3}
\begin{pmatrix}
\ell_1 & \ell_2 & \ell_3 \\
0 & 0 & 0 \\
\end{pmatrix}
^2 ,
\eeq
where $W_\ell$ is the mask power spectrum, and the last term is the Wigner-$3j$ symbol.
Then, we invert Eq.~\eqref{eq:pseudospec} to obtain the true power spectrum $C_\ell$
from the pseudo-power spectrum $\tilde{C}_\ell$.
Figure~\ref{fig:cl_mock} shows auto-power spectra of Compton-$y$
from 108 mock all-sky Compton-$y$ maps.
These mock measurements are used to estimate the covariance matrix
of tSZ auto-power spectra.
For a few realizations, the excessive signals can be seen.
These correspond to the cases where massive clusters are located
at low redshifts by chance.

\subsection{Mock shape catalog}
In order to create mock convergence maps,
we employ the mock shape catalog created in \citet{Shirasaki2019b}.
The mock catalog is specifically designed for the HSC survey and
constructed directly from the S16A shape catalog.
The same all-sky simulations \citep{Takahashi2017} in creating mock Compton-$y$ maps
are employed.
First, we randomly rotate the shapes of all galaxies in the catalog
to remove the lensing signal.
Then, we take the all-sky lensing map,
and deform the shape again according to the shear and convergence
at the position of each galaxy.
The redshift of each galaxy is determined by random sampling
from the PDF estimated with \texttt{MLZ}.
Finally, we obtain realistic catalogs which contain the lensing signal
and observational effects,
e.g., discrete distribution of source galaxies and
photometric redshift distribution.
Moreover, other quantities derived in the shape measurement,
e.g., lens weights, are also attached to the mock catalog and thus
we can carry out mock measurements in almost the same way to the real measurements.

\begin{figure*}
\includegraphics[width=\textwidth]{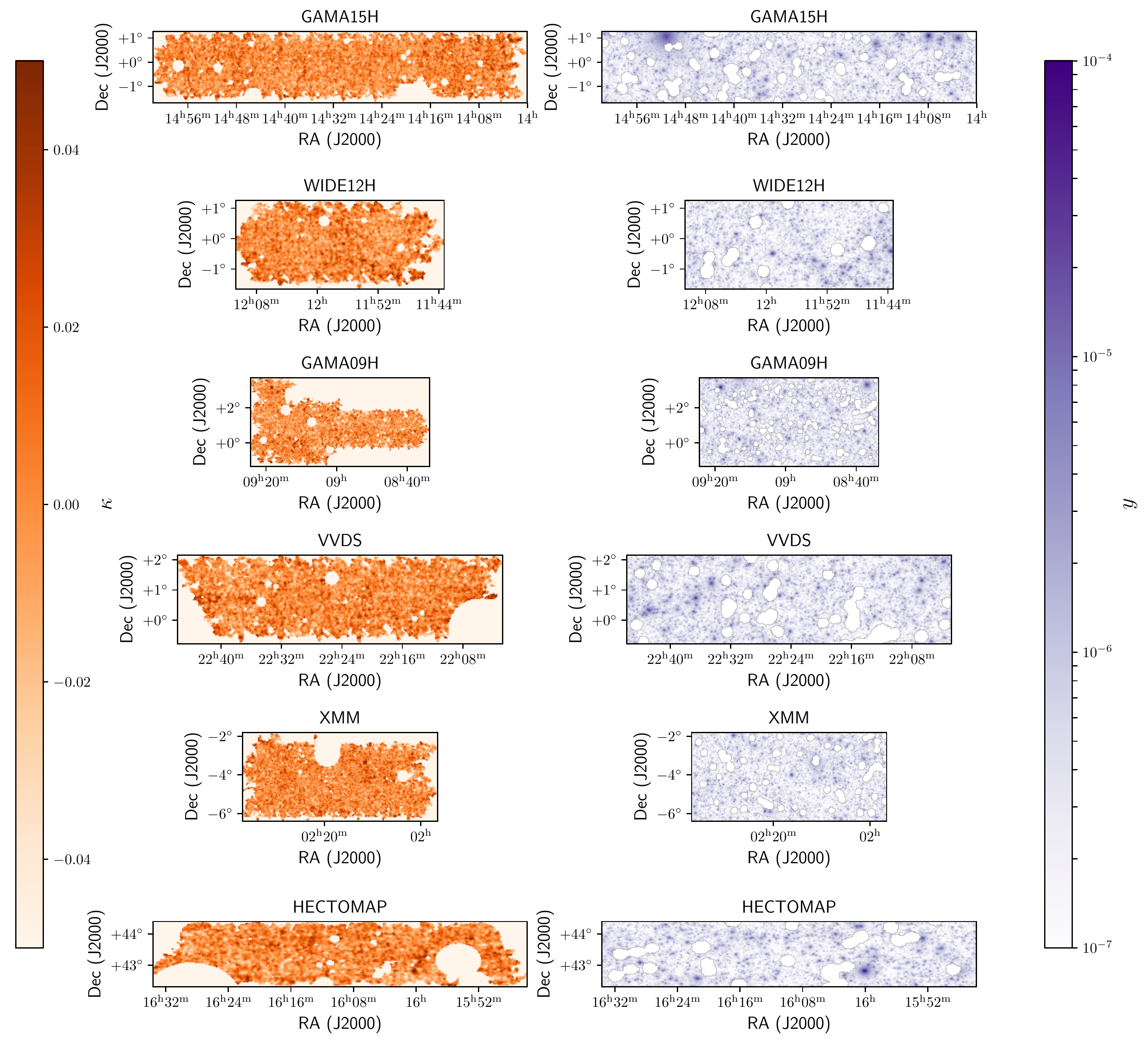}
\caption{An example of a mock convergence map (left panels)
and a mock Compton-$y$ map (right panels).
The Compton-$y$ maps are extracted from the original all-sky map
to show corresponding HSC S16A survey patches.}
\label{fig:mock}
\end{figure*}

\begin{figure}
\includegraphics[width=\columnwidth]{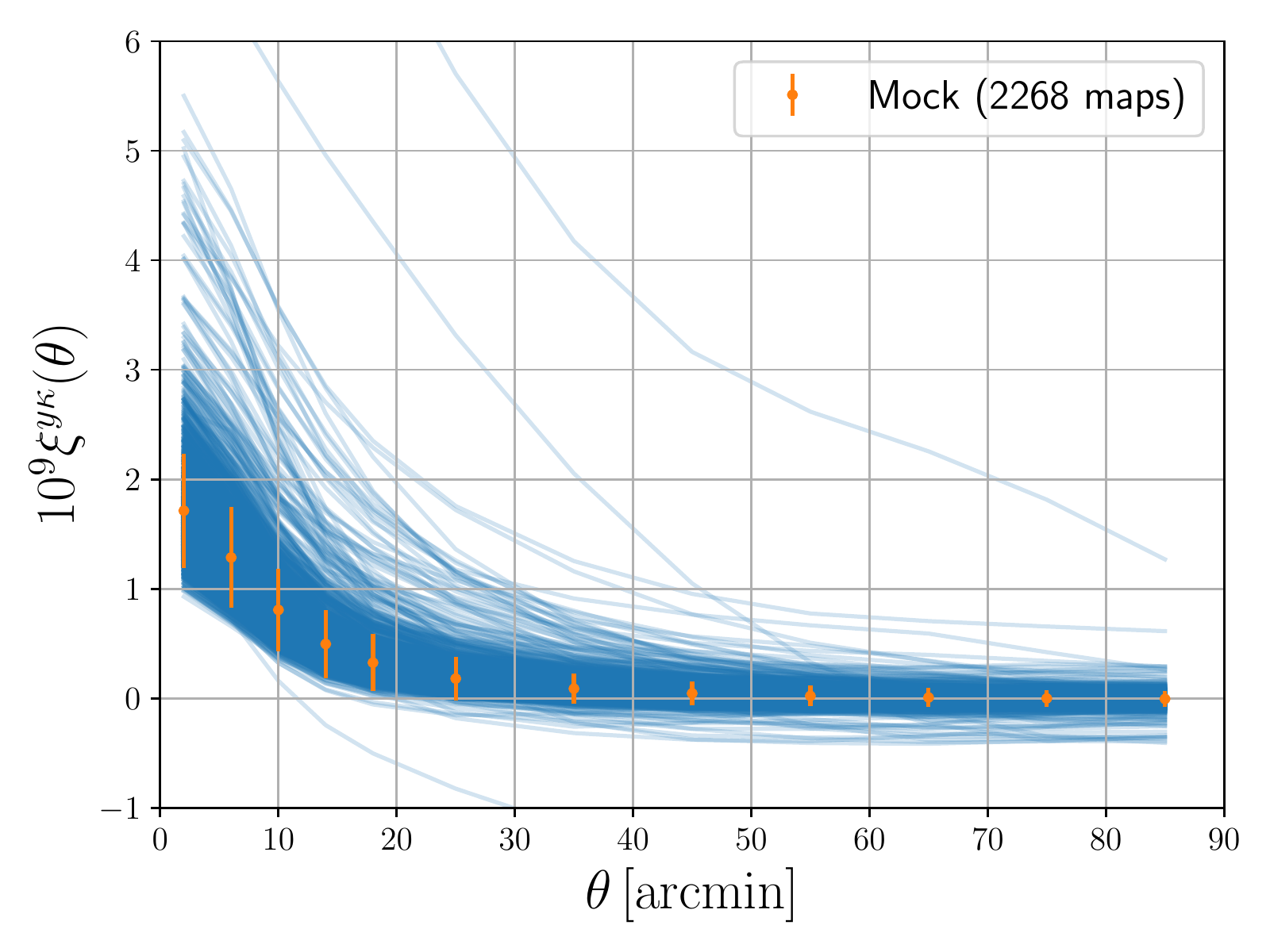}
\caption{The cross-correlations measured from mock Compton-$y$
and convergence maps.
Each blue solid line corresponds to one measurement with a mock map.
The orange points with error bars show
the mean and standard deviation among 2268 mock maps.}
\label{fig:xi_mock}
\end{figure}

Figure~\ref{fig:mock} shows convergence maps
and corresponding regions in the mock Compton-$y$ map.
It is clearly seen that massive clusters yield the strong signals
at the same positions in on both of the maps.
We can extract 21 HSC regions from one realization of the all-sky map,
and we compute the cross-correlation from these maps
for a total of $108 \times 21 = 2268$ realizations.
Figure~\ref{fig:xi_mock} shows cross-correlations obtained from our 2268 mock maps.
The mock measurements are used to estimate the covariance matrix of the cross-correlation.
Similarly to mock auto-power spectra, signals with very high amplitudes
are observed in several mock measurements.
This is also associated with massive clusters at low redshifts.

\section{Measurement}
\label{sec:measurement}
In this Section, we describe the analysis method of the cross-correlation
from the observational data from HSC S16A and \textit{Planck}.

\subsection{Cross-correlations of tSZ and WL}
\begin{figure}
\includegraphics[width=\columnwidth]{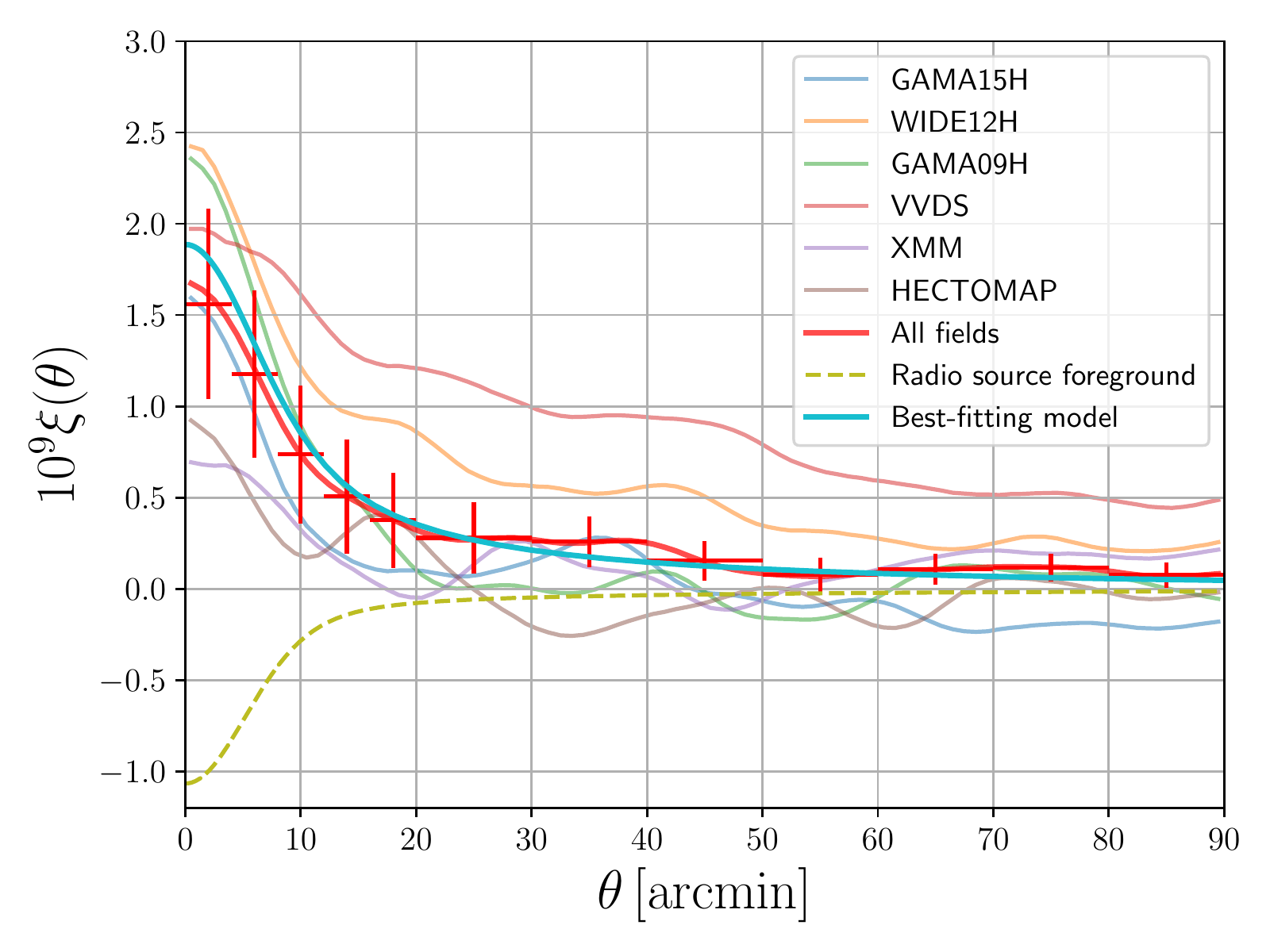}
\caption{The cross-correlations measured from the \textit{Planck}
Compton-$y$ map and HSC S16A convergence field.
The red error bars are estimated from mock measurements.
Each thin solid line corresponds to measurement in six different
HSC S16A survey patches.
The dashed line shows the radio foreground contribution
based on halo model in \citet{Shirasaki2019a}
with the best-fit parameter $B_\mathrm{R}$
and the solid cyan line shows the best-fitting halo model prediction,
where best-fit parameters are inferred from the cross only data set
with the \textit{Planck} prior (see Section~\ref{sec:full_space}).}
\label{fig:xi}
\end{figure}

We measure the tSZ-WL cross-correlation function with the following estimator:
\beqa
\hat{\xi}^{y \kappa} (\theta_k) &=&
\frac{\sum_{i, j} (\kappa(\bm{\vartheta}_i) -\kappa_0)
(y(\bm{\vartheta}_j) - y_0) w_{ij} (\theta_k)}{\sum_{i, j} w_{ij} (\theta_k)} , \\
w_{ij} (\theta_k) &\equiv&
\frac{1}{\sigma_\kappa^2 (\bm{\vartheta})} B(|\bm{\vartheta}_i-\bm{\vartheta}_j|, \theta_k)
\mathcal{W}_\kappa (\bm{\vartheta}_i) \mathcal{W}_y (\bm{\vartheta}_j) ,
\eeqa
where the subscript $k$ denotes the label of the angular bin.
The function $B(\vartheta ; \theta)$ denotes the binning scheme,
whose configuration is shown in Table~\ref{tab:cross_table}.
The quantities $\kappa_0$ and $y_0$ are the mean convergence computed for each patch
and mean Compton-$y$ in the \texttt{MILCA} Compton-$y$ map, respectively.
We apply the inverse variance weight $1/\sigma_\kappa^2 (\bm{\vartheta})$ for convergence
and equal weight for Compton-$y$.
The variance map for convergence field is estimated as follows.
First, we randomly rotate the ellipticity of galaxies in the shape catalog
and then reconstruct the convergence field with KS inversion.
We repeat this operation $300$ times and generate $300$ convergence maps.
The variance $\sigma_\kappa^2 (\bm{\vartheta})$ is computed as the sample variance
among these $300$ $E$-mode convergence maps.
It is also possible to apply the inverse variance weight for the Compton-$y$ map
because the variance map is also provided by the \textit{Planck} collaboration.
However, the high variance region has already been masked and both of the weighting
schemes yield consistent results.
Therefore, we adopt the equal weight for Compton-$y$.
The survey window functions $\mathcal{W}_\kappa (\bm{\theta})$ and
$\mathcal{W}_y (\bm{\theta})$ take zero
when the angular position $\bm{\theta}$ is masked and otherwise unity.
The mask of Compton-$y$ map is composed of $40\%$ galactic mask
and point source mask, and convergence field is masked
for the positions where
the smoothed galaxy number density is less than the half of the mean density in each patch.
We subtract the mean signal for convergence
because the KS inversion cannot reconstruct the uniform signal.
We also subtract the mean Compton-$y$ because the mean of the mock Compton-$y$ map
does not vanish by nature, but the subtraction does not affect the measurement with the real data
because the mean is already close to zero due to noise. The means are
computed as\footnote{When the inverse variance weight is introduced in the mean calculation,
the difference from the equal weight is negligible.}
\beqa
\kappa_0 &=& \frac{\int d^2 \theta \, \mathcal{W}_\kappa (\bm{\theta}) \kappa (\bm{\theta})}
{\int d^2 \theta \, \mathcal{W}_\kappa (\bm{\theta})} , \\
y_0 &=& \frac{\int d^2 \theta \, \mathcal{W}_y (\bm{\theta}) y(\bm{\theta})}
{\int d^2 \theta \, \mathcal{W}_y (\bm{\theta})} .
\eeqa
Instead of using cross-power spectrum directly,
we calculate the cross-correlations, in which we
can incorporate the masking effect
in a straightforward way.
To derive the prediction of the cross-correlation,
the Hankel transformation is applied to the cross-power spectrum based on halo model:
\beq
\xi^{y \kappa} (\theta) = \int \frac{\ell d\ell}{2 \pi}
C^{y \kappa} (\ell) \tilde{W}_\mathrm{G} (\ell ; \theta_s^\kappa)
\tilde{W}_\mathrm{G} (\ell ; \theta_s^y) J_0 (\ell \theta) ,
\eeq
where $\tilde{W}_\mathrm{G} (\ell ; \theta_s)$ is the Fourier transform
of the Gaussian window function:
\beq
\tilde{W}_\mathrm{G} (\ell ; \theta_s) = \exp \left( - \frac{1}{4} \ell^2 \theta_s^2 \right) .
\eeq
In Figure~\ref{fig:xi}, the measurement of the cross-correlation function
for each patch in HSC S16A and
the radio foreground contribution (see Section~\ref{sec:foreground}) are shown.
Note that the error bar corresponds to the standard deviation for the all HSC S16A regions.
That is why the measured cross-correlation for each patch looks statistically inconsistent
with the measurement for the all patches
but we confirm that the variances are within the statistical uncertainty
when the error are scaled according to the area of each patch.

\begin{table*}
\caption{The binning of angular separations, the cross-correlations
measured with \textit{Planck} and HSC S16A,
the standard deviation estimated from mock observations,
and templates for radio source contributions \citep{Shirasaki2019a}.}
\label{tab:cross_table}
\centering
\begin{tabular}{cccccc}
  $\theta_\mathrm{min} \, [\mathrm{arcmin}]$ & $\theta_\mathrm{max} \, [\mathrm{arcmin}]$
  & $\theta \, [\mathrm{arcmin}]$ & $10^{9} \xi^{y\kappa} (\theta)$ &
  $10^{9} \sigma^{y\kappa} (\theta)$ & $10^{9} \xi_\mathrm{R} (\theta)$ \\
  \hline
  \hline
  $0.0$ & $4.0$ & $2.0$ & $1.559815$ & $0.521110$ & $-0.775484$ \\
  $4.0$ & $8.0$ & $6.0$ & $1.175148$ & $0.458264$ & $-0.476844$ \\
  $8.0$ & $12.0$ & $10.0$ & $0.736128$ & $0.378507$ & $-0.220432$ \\
  $12.0$ & $16.0$ & $14.0$ & $0.505716$ & $0.312014$ & $-0.109852$ \\
  $16.0$ & $20.0$ & $18.0$ & $0.376050$ & $0.261536$ & $-0.070340$ \\
  $20.0$ & $30.0$ & $25.0$ & $0.279293$ & $0.196774$ & $-0.045258$ \\
  $30.0$ & $40.0$ & $35.0$ & $0.258154$ & $0.139731$ & $-0.031025$ \\
  $40.0$ & $50.0$ & $45.0$ & $0.153550$ & $0.108164$ & $-0.023614$ \\
  $50.0$ & $60.0$ & $55.0$ & $0.075866$ & $0.092662$ & $-0.018867$ \\
  $60.0$ & $70.0$ & $65.0$ & $0.106327$ & $0.085000$ & $-0.015500$ \\
  $70.0$ & $80.0$ & $75.0$ & $0.116000$ & $0.077772$ & $-0.012971$ \\
  $80.0$ & $90.0$ & $85.0$ & $0.075386$ & $0.069763$ & $-0.011004$ \\
  \hline
\end{tabular}
\end{table*}

\subsection{Null tests}
\label{sec:null_tests}
In order to confirm the cross-correlation signal is significant
and is not spurious due to systematic effects,
we measure the cross-correlation of Compton-$y$ map and
auxiliary maps of $B$-mode convergence, PSF leakage, and PSF residual.
All of the cross-correlations should be consistent
with null detections.
The $B$-mode convergence is obtained through the regular analysis
and it corresponds to the imaginary part of the convergence field
obtained by the KS inversion.
In the estimator of the cross-correlation for the $B$-mode map,
we need the inverse variance for each pixel as the weight.
The variance is estimated from 300 randomly rotated maps,
which are also used to estimate the variance in the $E$-mode case.
Then, the variance is computed from the imaginary part of convergence field
reconstructed from randomly rotated maps.
The shape catalog also provides the model estimate of
PSF ellipticity $e_\mathrm{p}$ at positions of source galaxies.
Then, we carry out the KS inversion
from the PSF estimates, and cross-correlate it with the Compton-$y$ map.
For PSF residual, we can obtain the true PSF from images of stars,
which are reserved for the PSF estimation \citep{Bosch2018},
and take the difference of the true PSF and model estimates,
$e_\mathrm{q} \equiv e_\mathrm{p} - e_\mathrm{star}$.
Similarly to PSF
ellipticity $e_\mathrm{p}$, we repeat KS inversion
and cross-correlation measurements with PSF residual $e_\mathrm{q}$.
For the cross-correlation measurements with PSF leakage and PSF residual,
we adopt equal weight in the estimator instead of inverse variance.

\begin{figure}
\includegraphics[width=0.9\columnwidth]{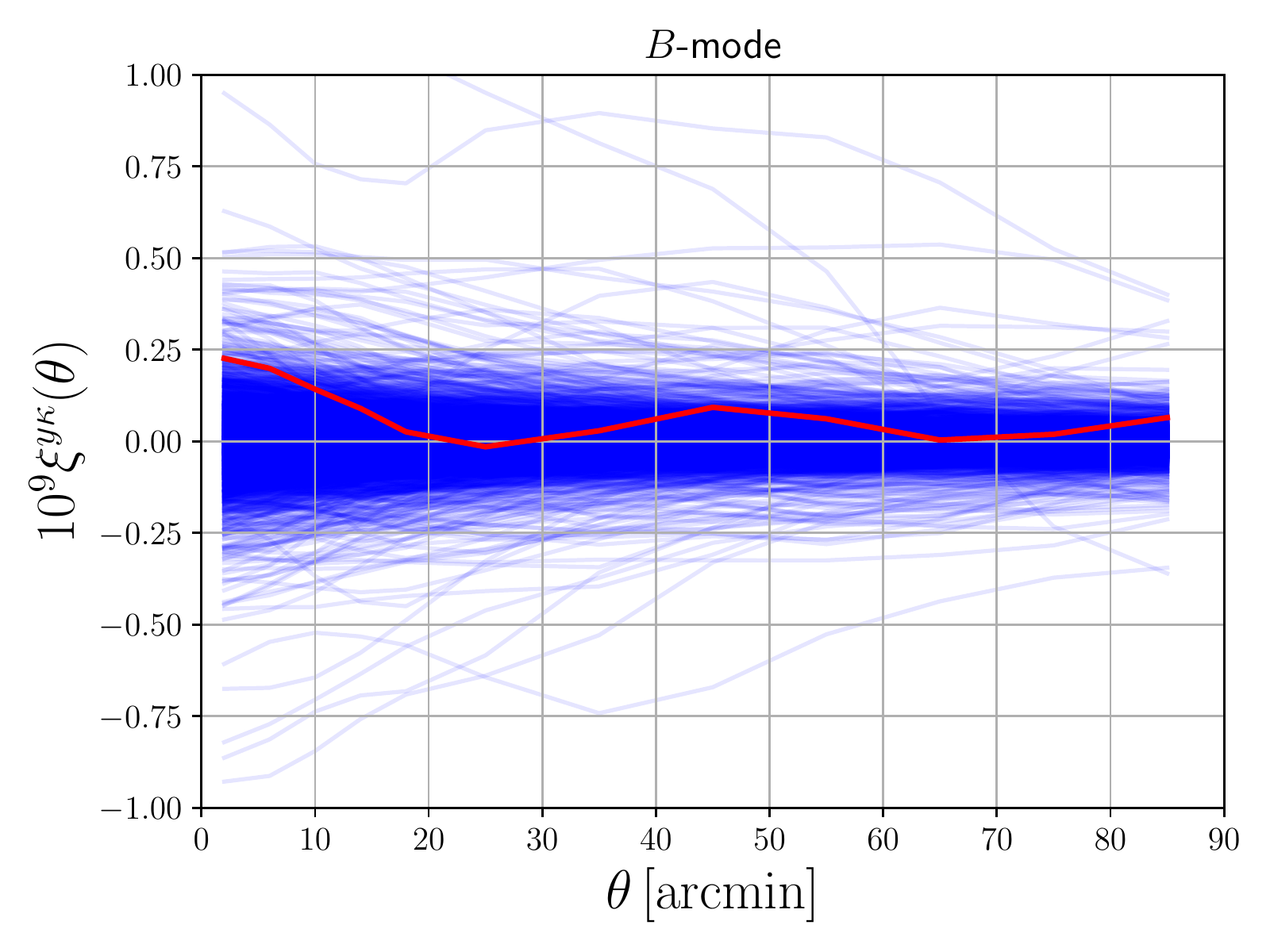}
\includegraphics[width=0.9\columnwidth]{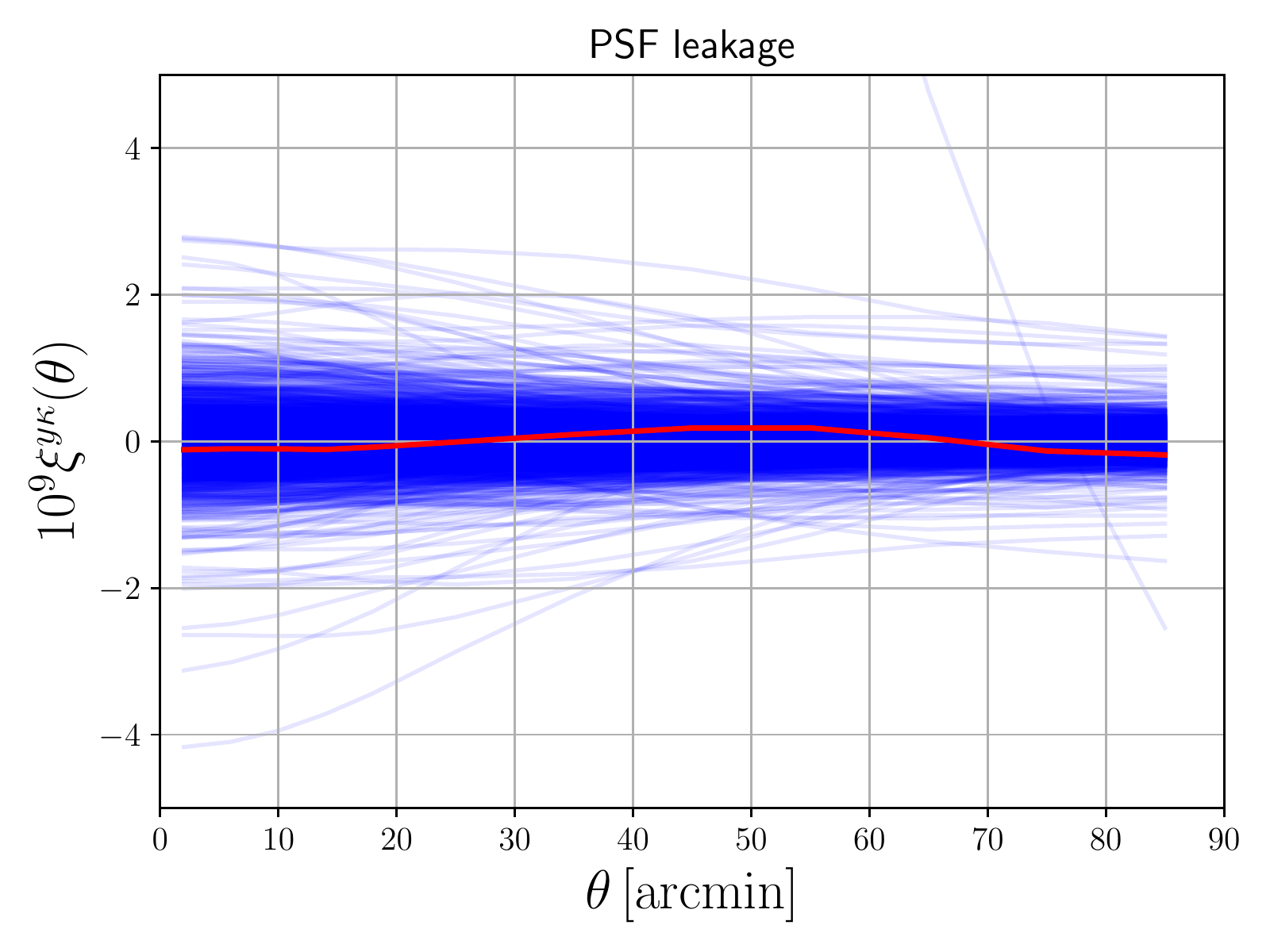}
\includegraphics[width=0.9\columnwidth]{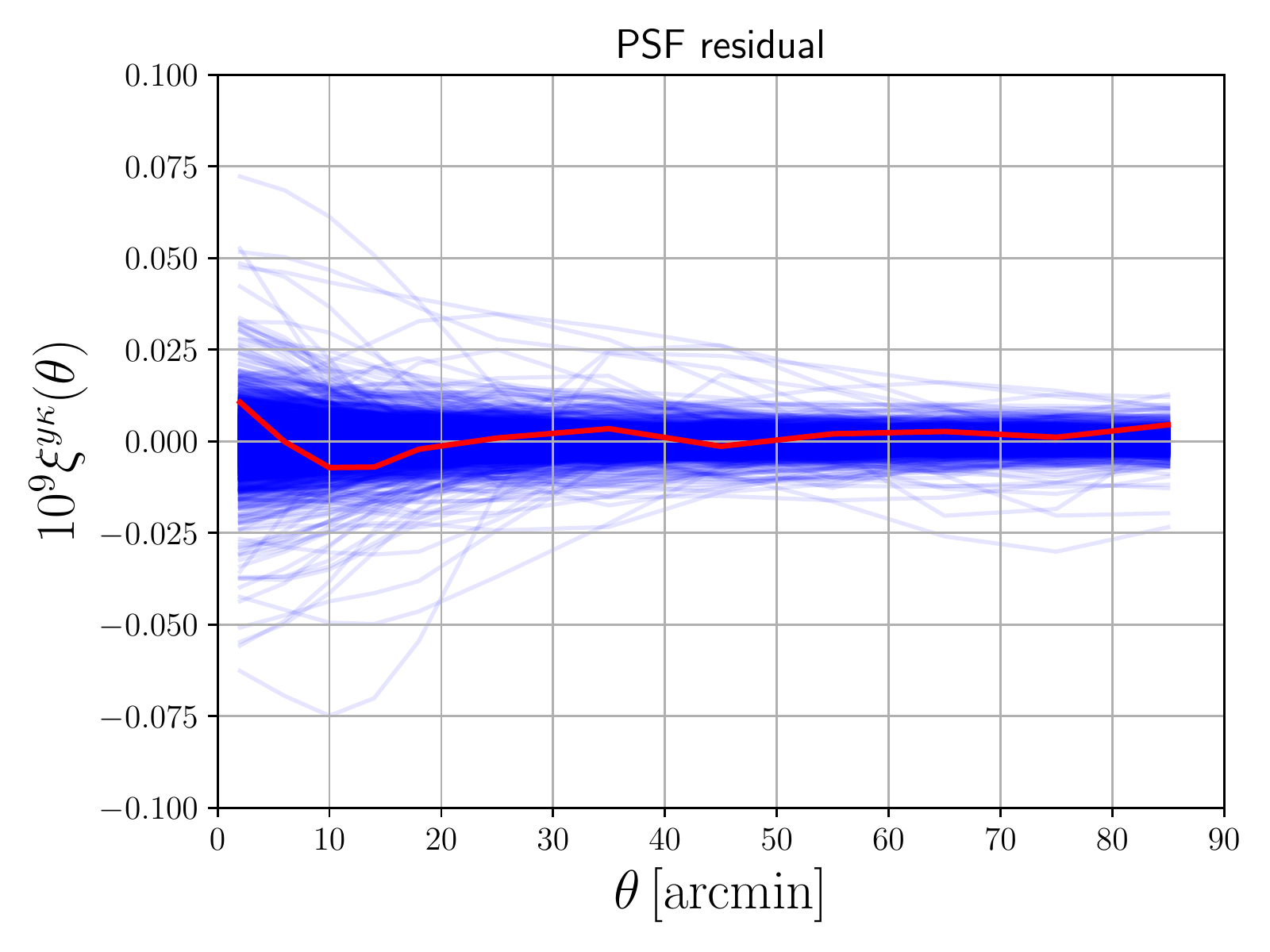}
\caption{The cross-correlations with $B$-mode, PSF leakage, and PSF residual
maps and real or mock Compton-$y$ maps.
The red line is the measurement with real data.
There are 2268 blue lines, each of which corresponds to the measurement with
one mock Compton-$y$ map.}
\label{fig:null_tests}
\end{figure}

In order to evaluate the statistical significance, i.e., $p$-value,
with respect to null signals,
we make use of mock Compton-$y$ maps again.
We measure the cross-correlations between $B$-mode, PSF leakage, and PSF residual map,
and mock Compton-$y$ map $2268$ times.
Since these auxiliary maps and mock maps should be uncorrelated,
null signals with statistical variance are expected.
Then, we compute the chi-square for each measurement:
\beq
\chi^2_r = \sum_{i,j} \xi_r (\theta_i) \mathrm{Cov}^{-1}_{i j}
\xi_r (\theta_j) \ (r = 1, \ldots, 2268),
\eeq
where the covariance matrix is estimated
from $2268$ mock measurements.\footnote{The inverse covariance matrix is estimated
with the method described in Appendix~\ref{sec:invcov}
because inversion of the high dimensional covariance matrix for correlated data
may lead to numerically unstable estimation.}
The $p$-value is defined as the number of mock measurements
which exceed the chi-square computed with real data.
Table~\ref{tab:p-values} shows $p$-values for $B$-mode,
PSF leakage, and PSF residual, and
we show the cross-correlation measurements
between mock Compton-$y$ maps and
convergence fields
from $B$-mode, PSF leakage, and PSF residual in Figure~\ref{fig:null_tests}.
The $p$-value for each null test is $4.14\%$ ($B$-mode),
$17.68\%$ (PSF leakage), and $4.10\%$ (PSF residual),
which corresponds to $1.73\sigma$ ($B$-mode), $0.93\sigma$ (PSF leakage),
and $1.74\sigma$ (PSF residual) for Gaussian distribution.
Thus, we can conclude that the measurements with these three maps
are consistent with null signals.

\begin{table}
\caption{The $p$-values for null tests of
$B$-mode, PSF leakage, and PSF residual,
which are estimated with mock measurements.
The corresponding variances for Gaussian distribution
are also shown.}
\label{tab:p-values}
\begin{center}
\begin{tabular}{c|c}
Map for null test & $p$-value \\
\hline
\hline
$B$-mode & $4.14\%$ ($1.73\sigma$) \\
PSF leakage & $17.68\%$ ($0.93\sigma$) \\
PSF residual & $4.10\%$ ($1.74\sigma$) \\
\hline
\end{tabular}
\end{center}
\end{table}

\subsection{Statistical significance of the measurement}
\label{sec:significance}
\begin{figure}
\includegraphics[width=\columnwidth]{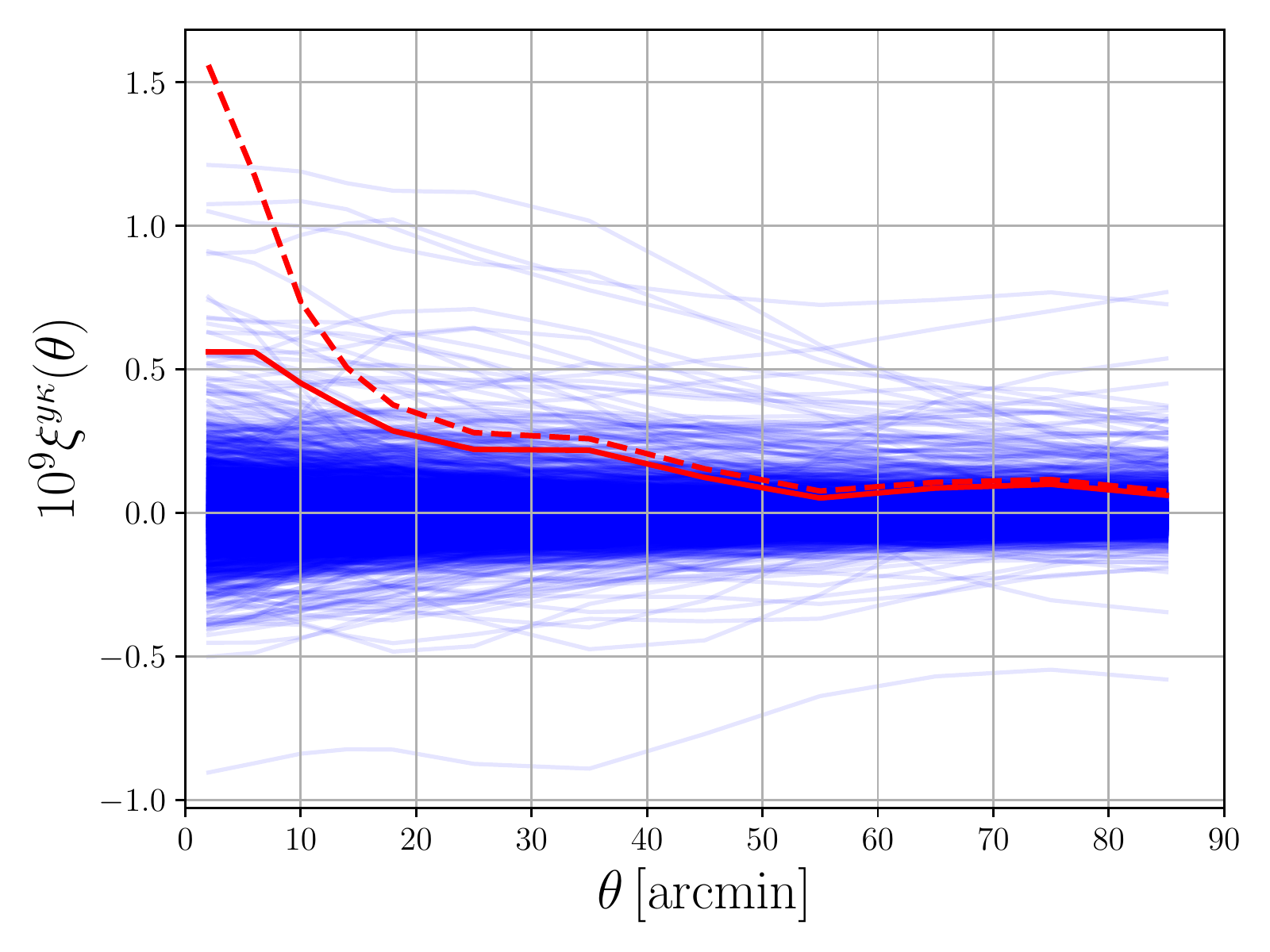}
\caption{The cross-correlations with $E$-mode map
and real or mock Compton-$y$ maps.
The red solid line is the real measurement with the radio contribution subtracted
with the best-fit parameter $B_\mathrm{R}$, which is inferred
from the cross only data set with the \textit{Planck} prior (see Appendix~\ref{sec:full_space}).
The real measurement before subtraction of the radio contribution
is also shown as the red dashed line.
There are 2268 blue lines, each of which corresponds to measurement with
one mock Compton-$y$ map.}
\label{fig:significance}
\end{figure}

Here, we evaluate the statistical significance of the $E$-mode signal.
Similarly to the null tests, we carry out
cross-correlation measurements with $E$-mode convergence and
mock Compton-$y$ maps, and compute chi-square for each measurement.
Figure~\ref{fig:significance} shows the real measurement with and without
foreground subtraction and mock measurements with real and mock Compton-$y$ maps.
It is expected that the mock measurement should be null because
we cross-correlate the real convergence map and the mock Compton-$y$ maps
in contrast to measurements with mock convergence maps and
mock Compton-$y$ maps (see Figure~\ref{fig:xi_mock}),
where the significant signal is expected.
The $p$-value corresponds to the fraction of mock measurements
which chi-square exceeds the one from the true measurement.
The derived $p$-value for the $E$-mode cross-correlation is $0.0441\%$,
which corresponds to $3.33\sigma$ for Gaussian distribution.
On the other hand, the $E$-mode cross-correlation contains
the contamination due to foreground radio emission.
We subtract the contribution
from the measured $E$-mode cross-correlation
with the best-fit amplitude parameter $B_\mathrm{R}$ (see Section~\ref{sec:foreground}
and Eq.~\ref{eq:xi_foreground})
inferred with the cross only data set
with the \textit{Planck} prior (see Section~\ref{sec:inference}).
After the foreground removal, we then recompute the chi-square
with respect to null signal.
The resultant $p$-value is $2.38\%$,
which corresponds to $1.98\sigma$ for Gaussian distribution.

\section{Cosmological analyses}
\label{sec:analyses}

\subsection{Foreground contribution}
\label{sec:foreground}
Due to imperfect separation of foreground components
in constructing Compton-$y$ map,
the measured auto-power spectrum and cross-correlations
contain the contribution from the foreground.
Such foregrounds include cosmic infrared background (CIB),
radio point sources, and infrared point sources.
In our analysis, we take into account these contributions
following \citet{Bolliet2018} for the tSZ auto-power spectrum
and \citet{Shirasaki2019a} for the tSZ-WL cross-correlation.

\begin{figure}
\includegraphics[width=\columnwidth]{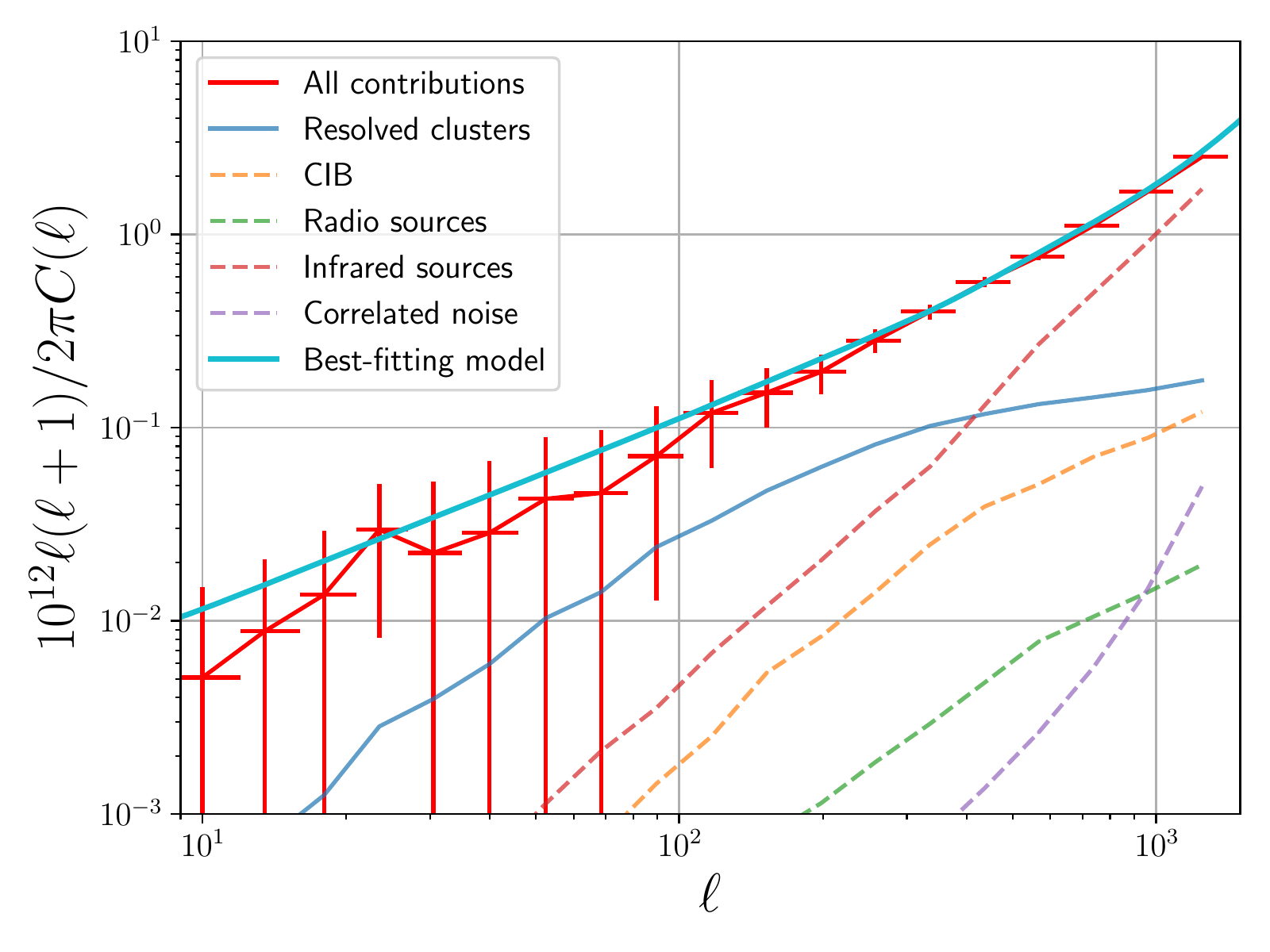}
\caption{The auto-power spectra of tSZ measured from \textit{Planck}
and templates for the foreground contributions of CIB, RS, and IR,
and the correlated noise contribution.
The amplitude of the correlated noise is $A_\mathrm{CN} = 0.903$.
The solid cyan line shows the best-fitting halo model prediction
and the dashed lines show the best-fitting foreground power spectra
where best-fit parameters are inferred from the auto only data set
with the \textit{Planck} prior (see Appendix~\ref{sec:full_space}).
The contribution from resolve clusters is also shown as the blue solid line.
The red error bars are estimated from mock measurements.}
\label{fig:cl}
\end{figure}

First, we briefly describe the foreground treatment in the auto-power spectrum
in \citet{Bolliet2018}. They consider three foreground contributions:
CIB, radio point sources (RS), and infrared point sources (IR).
In addition to these components, the residual correlated noise (CN) is also included.
The power spectrum templates for these components are given in \citet{Planck2015XXII}.
The amplitudes of the power spectra are modelled as nuisance parameters,
and in summary, the total power spectrum $\hat{C}^{yy}$ is given as
\beqa
\hat{C}^{yy} (\ell) &=& C^{yy} (\ell) + A_\mathrm{CIB} C_\mathrm{CIB} (\ell) +
A_\mathrm{IR} C_\mathrm{IR} (\ell) + \nonumber \\
&& A_\mathrm{RS} C_\mathrm{RS} (\ell) + A_\mathrm{CN} C_\mathrm{CN} (\ell) ,
\eeqa
where $C^{yy}$ is the prediction based on halo model,
$C_\mathrm{CIB}$, $C_\mathrm{IR}$, $C_\mathrm{RS}$,
and $C_\mathrm{CN}$ are templates for CIB, IR, RS, and CN, respectively.
The amplitude of CN $A_\mathrm{CN}$ is determined with the measurement
at the highest multipole ($\ell = 2742$)
because the CN term is dominant at small scales.
As a result, we fix the amplitude of the CN term as
$A_\mathrm{CN} = \hat{C}^{yy} (\ell = 2742) / C_\mathrm{CN} (\ell = 2742)
= 0.903$.
Other amplitudes ($A_\mathrm{CIB}$, $A_\mathrm{IR}$, and $A_\mathrm{RS}$)
are treated as nuisance parameters and marginalized in subsequent analysis.
The template power spectrum and total tSZ auto-power spectrum
is shown in Figure~\ref{fig:cl} and Table~\ref{tab:auto_table}.

\begin{table*}
\caption{The binning of multipoles, the auto-power spectrum measured by \textit{Planck},
the standard deviation estimated from mock observations,
the templates for CIB, RS, and IR, the contributions of CN and
resolved clusters (RC).
All of the template power spectra are based on \citet{Bolliet2018}.
Instead of the raw power spectrum $C(\ell)$,
the band-power $D (\ell) \equiv \ell (\ell+1)/(2 \pi) C(\ell)$ is shown.
Note that the largest multipole bin ($\ell = 1247.5$)
is used only in Eq.~\eqref{eq:RC_condition}.}
\label{tab:auto_table}
\begin{tabular}{cccccccccc}
  $\ell_\mathrm{min}$ & $\ell_\mathrm{max}$ & $\ell$ &
  $10^{12} D^{yy} (\ell)$ & $10^{12} \sigma^{yy} (\ell)$ & $10^{12} D_\mathrm{RC} (\ell)$ &
  $10^{12} D_\mathrm{CIB} (\ell)$ & $10^{12} D_\mathrm{RS} (\ell)$ & $10^{12} D_\mathrm{IR} (\ell)$ &
  $10^{12} D_\mathrm{CN} (\ell)$ \\
  \hline
  \hline
  $9$ & $12$ & $10.0$ & $0.005080$ & $0.009812$ & $0.000421$ & $0.000000$ & $0.000043$ & $0.000007$ & $0.000001$ \\
  $12$ & $16$ & $13.5$ & $0.008810$ & $0.011924$ & $0.000710$ & $0.000000$ & $0.000142$ & $0.000024$ & $0.000001$ \\
  $16$ & $21$ & $18.0$ & $0.013630$ & $0.015646$ & $0.001251$ & $0.000000$ & $0.000296$ & $0.000048$ & $0.000002$ \\
  $21$ & $27$ & $23.5$ & $0.029610$ & $0.021492$ & $0.002837$ & $0.000000$ & $0.000400$ & $0.000073$ & $0.000004$ \\
  $27$ & $35$ & $30.5$ & $0.022410$ & $0.030094$ & $0.003933$ & $0.000902$ & $0.000541$ & $0.000111$ & $0.000006$ \\
  $35$ & $46$ & $40.0$ & $0.028490$ & $0.038872$ & $0.005969$ & $0.002010$ & $0.001056$ & $0.000224$ & $0.000010$ \\
  $46$ & $60$ & $52.5$ & $0.042760$ & $0.046322$ & $0.010318$ & $0.003119$ & $0.001647$ & $0.000449$ & $0.000018$ \\
  $60$ & $78$ & $68.5$ & $0.045800$ & $0.051503$ & $0.014045$ & $0.006278$ & $0.002787$ & $0.000837$ & $0.000030$ \\
  $78$ & $102$ & $89.5$ & $0.071040$ & $0.058284$ & $0.024061$ & $0.012242$ & $0.004306$ & $0.001400$ & $0.000052$ \\
  $102$ & $133$ & $117.0$ & $0.119140$ & $0.057526$ & $0.032976$ & $0.021584$ & $0.006842$ & $0.002701$ & $0.000089$ \\
  $133$ & $173$ & $152.5$ & $0.151500$ & $0.051520$ & $0.047100$ & $0.045915$ & $0.011264$ & $0.004721$ & $0.000153$ \\
  $173$ & $224$ & $198.0$ & $0.193900$ & $0.045036$ & $0.062380$ & $0.070582$ & $0.016744$ & $0.008115$ & $0.000262$ \\
  $224$ & $292$ & $257.5$ & $0.281750$ & $0.039697$ & $0.081730$ & $0.119786$ & $0.027345$ & $0.014618$ & $0.000456$ \\
  $292$ & $380$ & $335.5$ & $0.398370$ & $0.036168$ & $0.101911$ & $0.211686$ & $0.043275$ & $0.024893$ & $0.000815$ \\
  $380$ & $494$ & $436.5$ & $0.567430$ & $0.033381$ & $0.117412$ & $0.332863$ & $0.070587$ & $0.051570$ & $0.001503$ \\
  $494$ & $642$ & $567.5$ & $0.768660$ & $0.032938$ & $0.132234$ & $0.434931$ & $0.115356$ & $0.107293$ & $0.002934$ \\
  $642$ & $835$ & $738.0$ & $1.110100$ & $0.031719$ & $0.143214$ & $0.602030$ & $0.154926$ & $0.197053$ & $0.006334$ \\
  $835$ & $1085$ & $959.5$ & $1.661400$ & $0.027203$ & $0.156202$ & $0.754733$ & $0.207200$ & $0.361713$ & $0.016171$ \\
  \hline
  $1085$ & $1411$ & $1247.5$ & $2.521700$ & $0.030189$ & $0.175341$ & $1.029014$ & $0.287652$ & $0.681036$ & $0.054883$ \\
  \hline
\end{tabular}
\end{table*}

Next, we discuss the foreground contribution in the tSZ-WL cross-correlations.
We employ the halo model prescription proposed in \citet{Shirasaki2019a},
where the template cross-power spectrum from extragalactic radio sources:
flat-spectrum radio quasars, BL Lac objects, and steep-spectrum sources.
\citet{Shirasaki2019a} also addresses the contribution from CIB,
but we do not include the contribution because the effect due to CIB is
subdominant for the cross-correlation.
The template is computed with respect to the radio frequency,
and to convert them into Compton-$y$, we employ the weight of Map C
in Table~1 of \citet{VanWaerbeke2014}.
The template is shown in Table~\ref{tab:cross_table}.
Since the weighting scheme is different for the \texttt{MILCA} Compton-$y$ map,
we introduce the amplitude parameter $B_\mathrm{R}$
and treat it as a nuisance parameter in the analysis.
As a result, the total cross-correlation $\hat{\xi}^{y \kappa}$ is given as
\beq
\label{eq:xi_foreground}
\hat{\xi}^{y \kappa} (\theta) = \xi^{y \kappa} (\theta) +
B_\mathrm{R} \xi_\mathrm{R} (\theta) ,
\eeq
where $\xi^{y \kappa} (\theta)$ is the predicted cross-correlation based on halo model.

\subsection{Inference of parameters}
\label{sec:inference}
In the analysis, we use tSZ-WL cross-correlations from HSC and \textit{Planck}
and tSZ auto-power spectrum from \textit{Planck} in order to
constrain cosmological parameters and the hydrostatic bias parameter $b_\mathrm{HSE}$.

\begin{figure}
\includegraphics[width=\columnwidth]{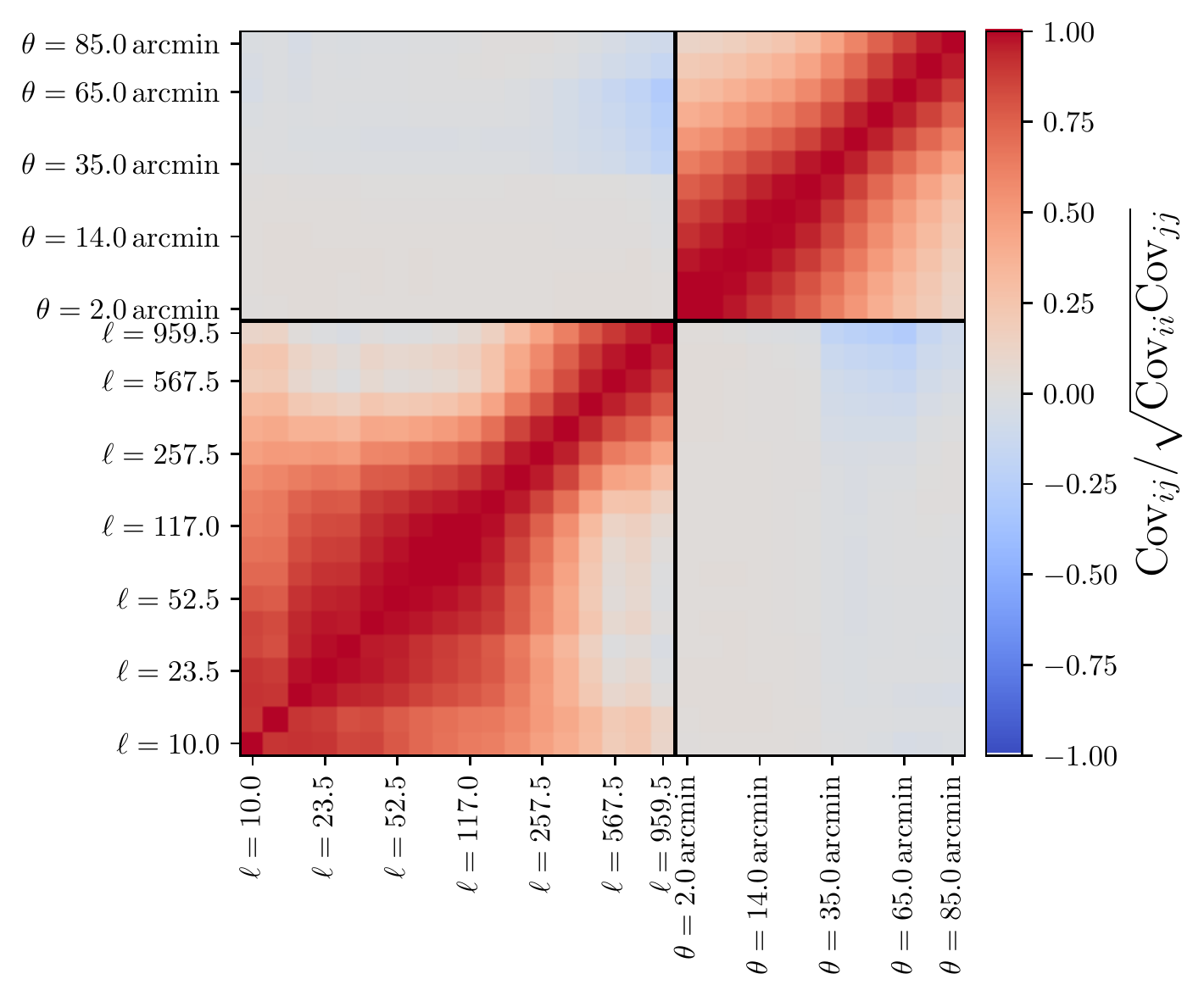}
\caption{Covariance matrix of the tSZ auto-power spectrum
and the tSZ-WL cross-correlation estimated from mock observations.}
\label{fig:cov}
\end{figure}

First, we define the data vector as
\beqa
D_C &=& (C^{yy} (\ell_1), \ldots, C^{yy} (\ell_{n_C}) ), \\
D_\xi &=& (\xi^{y\kappa} (\theta_1), \ldots, \xi^{y\kappa} (\theta_{n_\xi})), \\
D_{C+\xi} &=& (D_C, D_\xi) ,
\eeqa
where $n_C = 18$ and $n_\xi = 12$ are the number of bins
in auto-power spectrum and cross-correlation, respectively.
The binning is shown in Table~\ref{tab:auto_table} for the auto-power spectrum
and Table~\ref{tab:cross_table} for the cross-correlations.
Note that the largest multipole bin $\ell = 1247.5$ in the auto-power spectrum is
used only for the condition Eq.~\eqref{eq:RC_condition}.
From the mock measurements, we estimate covariance matrix,
\beq
\mathrm{Cov}_{ij} = \frac{1}{R-1} \sum_{r=1}^{R} (D^r_i - \bar{D}_i) (D^r_j - \bar{D}_j) ,
\eeq
where $R$ is the number of realizations, $\bar{D}$ is the sample mean of $R$ measurements,
and $r = 1, \ldots , R$ denotes the label of the realization.
There are $R = 108$ maps for mock Compton-$y$ and $R = 2268$ maps for mock convergence.
As a result, we have 108 and 2268 measurements of auto-power spectrum and
cross-correlations, respectively.
For estimation of cross-covariance between auto-power spectrum and cross-correlation,
we additionally generate $2268$ maps for Compton-$y$ by rotating the coordinates to
adjust the one in mock convergence map.
These additional mock Compton-$y$ maps are used for
cross-covariance estimation, null tests, and
evaluation of significance of cross-correlations.
The estimated covariance matrix of the tSZ auto-power spectrum
and the tSZ-WL cross-correlation is shown in Figure~\ref{fig:cov}.

The likelihood $\mathcal{L}$ is assumed to be multivariate Gaussian as
\beq
\log \mathcal{L} (\bm{p}) = -\frac{1}{2} \sum_{i,j} (\hat{D}_i - D_i (\bm{p}) )
\mathrm{Cov}^{-1}_{ij} (\hat{D}_j - D_j (\bm{p}) ) + \text{const.},
\eeq
where $\hat{D}$ is the measurement, $D (\bm{p})$ is the prediction based on the halo model,
and $\bm{p}$ is the parameter vector,
which includes cosmological parameters,
hydrostatic bias parameter $b_\mathrm{HSE}$, and nuisance parameters.
The cosmological parameters are composed of physical baryon density
$\omega_\mathrm{b} \equiv \Omega_\mathrm{b} h^2$,
physical CDM density $\omega_\mathrm{cdm} \equiv \Omega_\mathrm{cdm} h^2$,
scaled Hubble parameter
$h \equiv H_0/ (100 \, \mathrm{km} \, \mathrm{s}^{-1} \, \mathrm{Mpc}^{-1})$,
tilt and amplitude of
the scalar perturbation $n_\mathrm{s}$ and $\ln (10^{10} A_\mathrm{s})$.
In addition, we also consider three derived parameters: the total matter density
with respect to critical density, $\Omega_\mathrm{m}$, the amplitude
of the matter fluctuation at the scale of $8 \, h^{-1} \, \mathrm{Mpc}$, $\sigma_8$,
and the amplitude parameter $S_8 \equiv \sigma_8 (\Omega_\mathrm{m}/0.3)^{0.5}$,
which roughly corresponds to the amplitude of
cosmic shear power spectrum or correlation function.
Throughout the analysis, we assume the flat $\Lambda$CDM Universe
and there are three species of neutrinos,
one of which has finite mass of $m_\nu = 0.06 \, \mathrm{eV}$.
The total matter density with respect to critical density $\Omega_\mathrm{m}$ is
the sum of CDM $\Omega_\mathrm{cdm}$, baryon $\Omega_\mathrm{b}$, and
massive neutrinos $\Omega_{\nu} = m_\nu/(93.14 \, h^2 \,\mathrm{eV})$.
In addition to cosmological parameters,
we take the hydrostatic bias parameter $b_\mathrm{HSE}$ into account.
Nuisance parameters are introduced depending on data sets:
for the auto-power spectrum, amplitude parameters of
foreground contributions, $A_\mathrm{CIB}$, $A_\mathrm{IR}$, and $A_\mathrm{RS}$,
and for the cross-correlation, an amplitude parameter of radio foreground $B_\mathrm{R}$.
The details on how to estimate the inverse covarince matrix from
the sample covariance matrix are found in Appendix~\ref{sec:invcov}.
Only with auto-power spectrum and cross-correlations,
the constraining power is weak and it is hard to obtain converged results.
Therefore, we add priors on cosmological parameters from external measurements.
In this analysis, we consider two priors.
One is a combination of results of large-scale structure measurements
(hereafter LSS prior):
Joint Light-curve Analysis (JLA) of SDSS-II and SNLS for type Ia supernovae \citep{Betoule2014},
Baryon Oscillation Spectroscopic Survey (BOSS) Data Release 12
for baryon acoustic oscillations and
redshift space distortions \citep{Alam2017},
and HSC S16A WL cosmic shear power spectrum
\citep{Hikage2019}.\footnote{Though neutrinos are assumed to be massless
in the fiducial analysis of \citet{Hikage2019},
we use another chain, where the sum of neutrinos
is set to be $0.06 \, \mathrm{eV}$.}
The other one is the \textit{Planck} 2018 results in
TT,TE,EE+lowE+lensing dataset \citep{Planck2018VI,Planck2018VIII}
(hereafter \textit{Planck} prior).
The LSS prior includes additional four nuisance parameters (see Appendix~\ref{sec:full_space}).

Hence, the posterior distribution $\mathcal{P}$ is given as
\beq
\log \mathcal{P} (\bm{p}) = \log \mathcal{L} (\bm{p}) + \log P (\bm{p}) + \text{const}.
\eeq
We utilize the Markov chain
Monte-Carlo code \texttt{MontePython-3} \citep{Audren2013,Brinckmann2018}
to obtain chains for the posterior distribution.
In order to confirm the convergence of obtained chains,
we compute Gelman--Rubin statistic $R$ for all parameters
and run the analysis until the condition $R-1 < 0.01$ is reached.
For priors from HSC S16A cosmic shear power spectrum analysis or
\textit{Planck} 2018 results, we assume the multivariate Gaussian form as
\beq
\log P(\bm{p}) = -\frac{1}{2} \sum_{\alpha, \beta}
(p_\alpha - \bar{p}_\alpha) \mathcal{C}^{-1}_{\alpha \beta}
(p_\beta - \bar{p}_\beta) + \text{const.},
\eeq
where the mean $\bar{\bm{p}}$ and the covariance matrix $\mathcal{C}_{\alpha \beta}$
are estimated from parameter chains.\footnote{The inverse covariance
matrix $\mathcal{C}^{-1}_{\alpha \beta}$ is obtained
by simply inverting the sample covariance matrix $\mathcal{C}_{\alpha \beta}$
in contrast to the method described in Appendix~\ref{sec:invcov}
because the dimension of the matrix is not large.}
Note that the deviation from the multivariate Gaussian affects the resultant constraints
compared with the one with the full likelihood analysis.
For example, in the case with cosmic shear power spectrum,
the constraint on the amplitude parameter $S_8$ is significantly degraded
because the approximation cannot fully capture the shape of
the degeneracy between $\sigma_8$ and $\Omega_\mathrm{m}$.
However, it provides the reasonable estimates on the prior of most of parameters.
For other data sets (JLA and BOSS), we make use of likelihood packages provided
in \texttt{MontePython-3}.
For the prior on the hydrostatic bias parameter,
we adopt a hard prior $b_\mathrm{HSE} < 1$, which ensures
the hydrostatic mass is positive.
In addition, we impose an additional condition on
the auto-power spectrum following \citet{Bolliet2018}.
The contribution from the galaxy clusters which have
already been observed in X-ray or SZ has been measured
and is shown in Table~\ref{tab:auto_table}.
The prediction should exceed the contribution at least small
scales, where the absolute errors are small, i.e.,
\beq
\label{eq:RC_condition}
C^{yy} (\ell) - C_\mathrm{RC} (\ell) > 0.
\eeq
This condition should be satisfied for all multipoles between $\ell = 257.5$
and $\ell = 1247.5$.
The power spectrum at large scales has large statistical variances
and is not subject to the condition. If this condition is not satisfied,
we force posterior to be zero.

\begin{figure*}
\includegraphics[width=\textwidth]{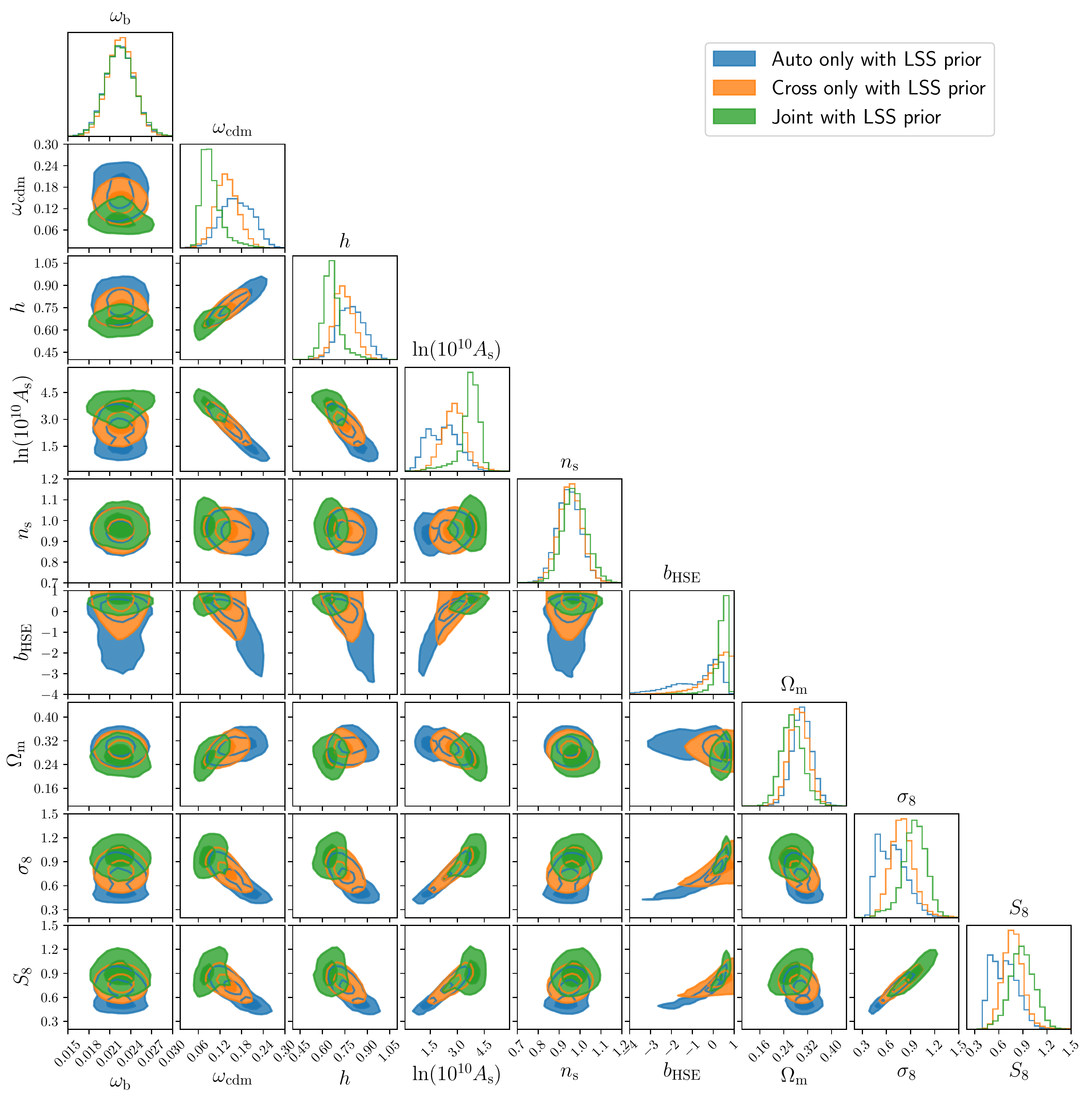}
\caption{Confidence regions of cosmological parameters and hydrostatic bias parameter
inferred with three data sets (auto only, cross only, and joint)
with the LSS prior.
The inner (outer) contour corresponds to the $1\sigma$ ($2\sigma$) confidence level.
The diagonal panels show the marginalized likelihoods.}
\label{fig:triangle_LSS}
\end{figure*}

\begin{figure*}
\includegraphics[width=\textwidth]{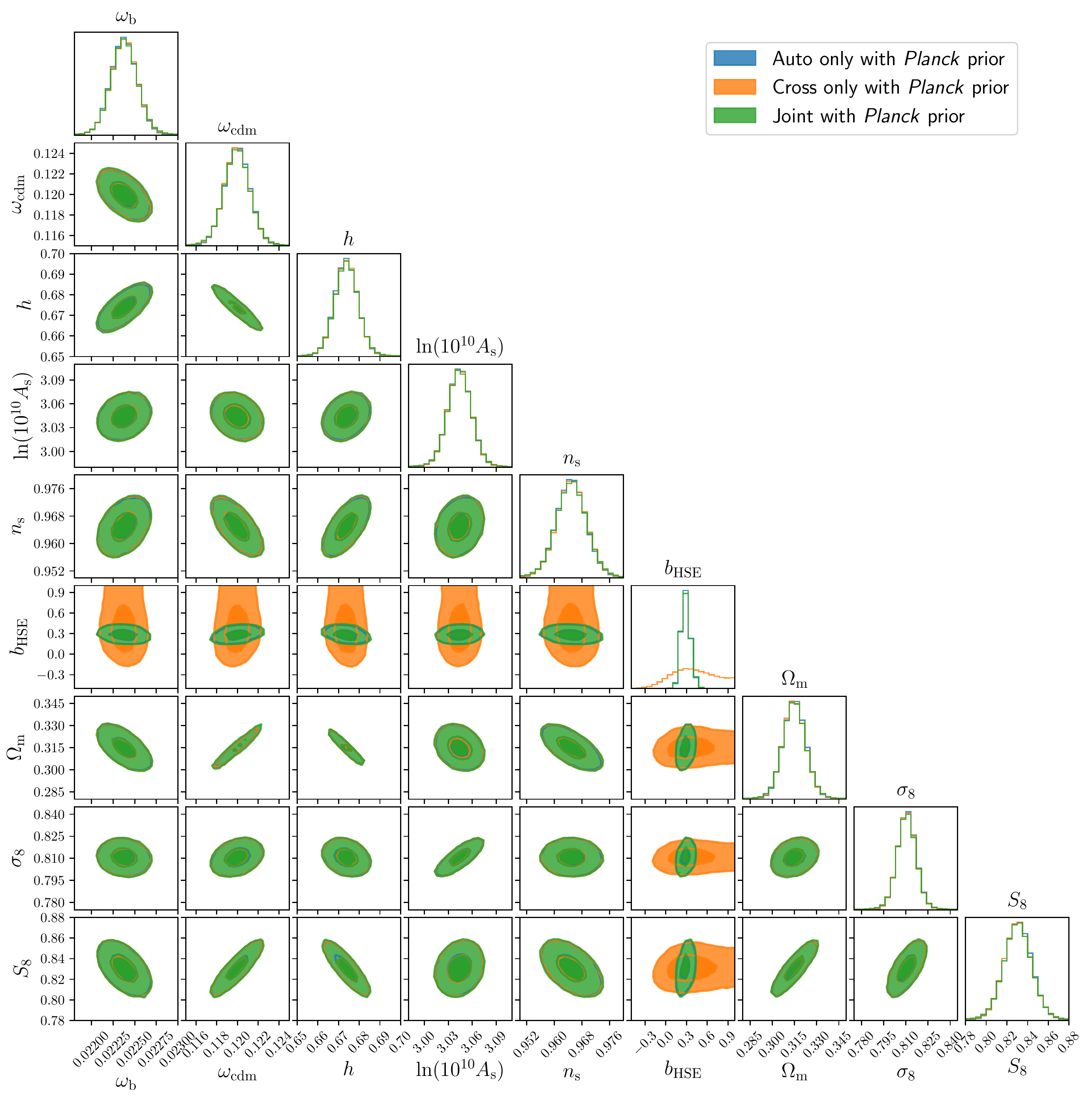}
\caption{Confidence regions of cosmological parameters and hydrostatic bias parameter
inferred with three data sets (auto only, cross only, and joint)
with the \textit{Planck} prior.
The inner (outer) contour corresponds to the $1\sigma$ ($2\sigma$) level.
The diagonal panels show the marginalized likelihoods.}
\label{fig:triangle_Plancklensing}
\end{figure*}

We show constraints of parameters with tSZ auto-power spectrum
and tSZ-WL cross-correlations for LSS and \textit{Planck} priors
in Figures~\ref{fig:triangle_LSS} and \ref{fig:triangle_Plancklensing}, respectively.
The results in the full parameter space (cosmological parameters, hydrostatic bias,
nuisance parameters, and derived parameters) are found in Appendix~\ref{sec:full_space}.
In each case, we present constraints with three data sets:
tSZ auto-power spectrum only (hereafter, auto only),
tSZ-WL cross-correlation only (cross only),
and joint analysis with both of them (joint).
Since $b_\mathrm{HSE}$
is highly degenerated with cosmological parameters,
the tighter constraints can be obtained with \textit{Planck} prior because
this prior determines the cosmological parameters better.
As a whole, the constraining power of tSZ auto-power spectrum is better than
that of tSZ-WL cross-correlations due to the large survey area.
Though errors are large for the case of the cross-correlation only data set,
all three data sets give consistent results for all cosmological parameters and $b_\mathrm{HSE}$.
The constraints with the joint data set do not necessarily lie between the results
with the auto only and cross only data sets because of the cross-covariance between
the auto-power spectrum and cross-correlation.

\begin{figure}
\includegraphics[width=\columnwidth]{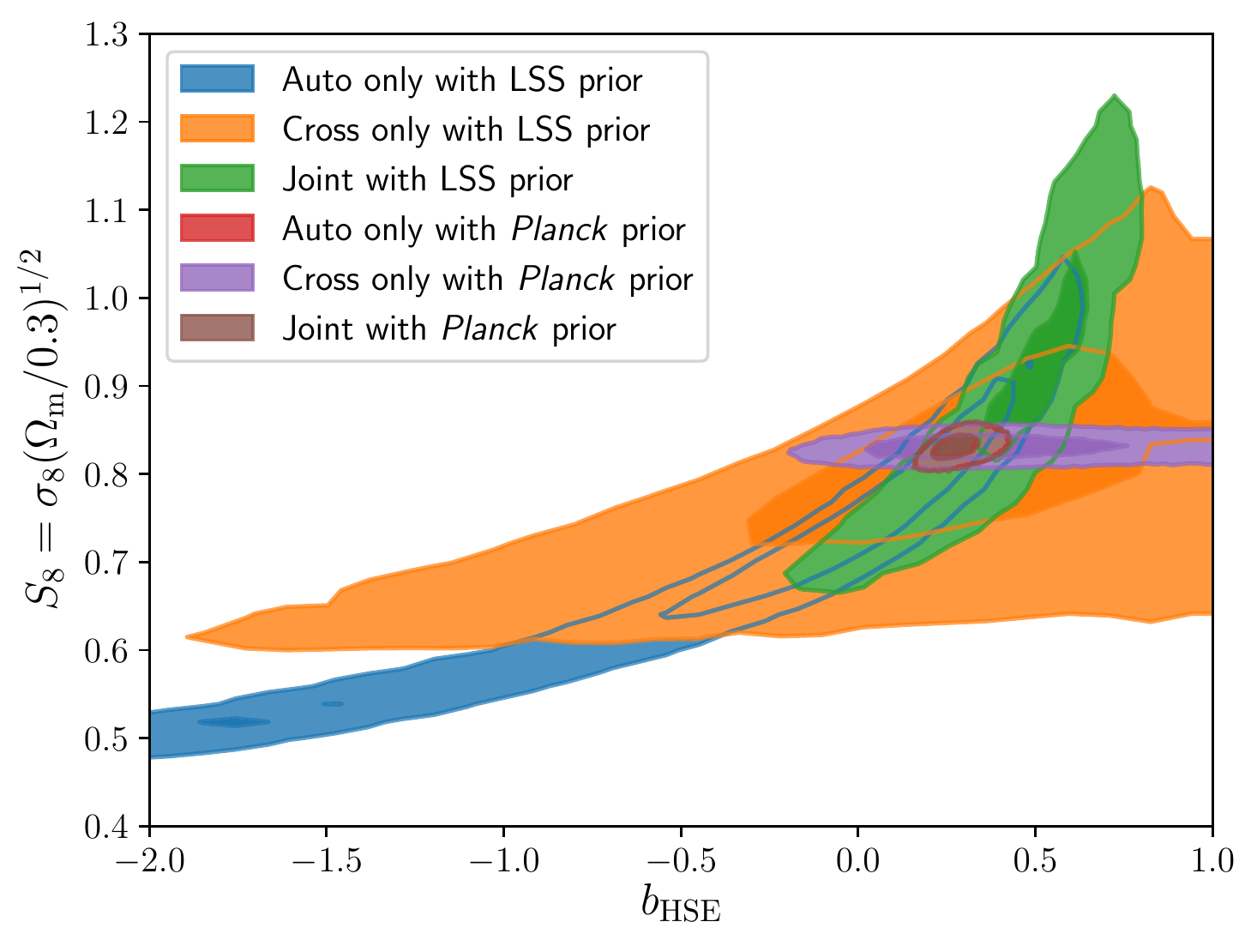}
\caption{Confidence regions of hydrostatic bias and amplitude parameter $S_8$
for three data sets with two priors.
The inner (outer) contour corresponds to the $1\sigma$ ($2\sigma$) level.}
\label{fig:b_S8}
\end{figure}

In Figure~\ref{fig:b_S8}, the constraints
on the amplitude parameter $S_8$
and the hydrostatic bias parameter $b_\mathrm{HSE}$ are shown.
Due to weak constraining power on cosmological parameters of the LSS prior,
the error contour for data sets with the prior is much larger than
those with the \textit{Planck} prior,
but all of results are consistent with each other
and the tightest constraint, which is obtained with auto only or joint data sets
with the \textit{Planck} prior, prefers the amplitude parameter $S_8 \simeq 0.83$
and the hydrostatic bias parameter $b_\mathrm{HSE} \simeq 0.3$.

\section{Discussions}
\label{sec:discussions}

\subsection{Contributions from resolved clusters}
\begin{figure}
\includegraphics[width=\columnwidth]{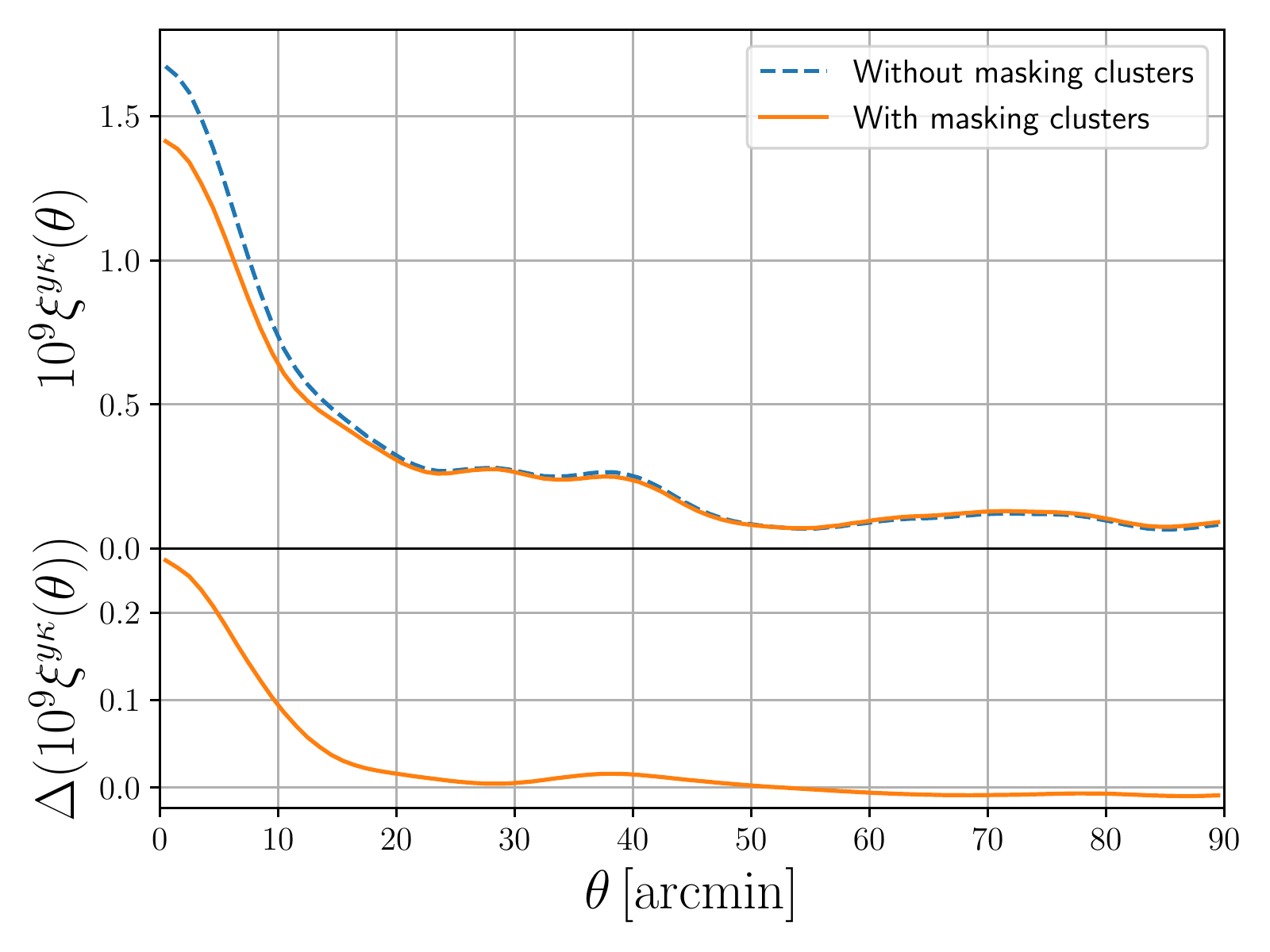}
\caption{The cross-correlations with (solid line) and without (dashed line)
SZ-detected clusters in HSC S16A survey footprints masked \citep{Medezinski2018}.
The lower panel shows the difference between these two measurements.}
\label{fig:xi_RC}
\end{figure}

In order to estimate the contributions from clusters
which have already been detected both through HSC WL surveys and \textit{Planck}
SZ observations,
we repeat the measurement with the additional mask covering
the detected clusters.
We use the cluster catalog in \citet{Medezinski2018},
which contains $5$ clusters located within the HSC S16A footprints
and the SZ signals have already been detected by \textit{Planck} \citep{Planck2015XXVII}.
The locations, redshifts, and masses which are inferred by fitting the WL signal
with the NFW profile are shown in Table~\ref{tab:cl_cat}.
We mask the regions within the angular extent $\theta_{200} = R_{200} / d_A (z)$
for each cluster,
where $d_A (z)$ is the angular diameter distance.
In Figure~\ref{fig:xi_RC}, we show the cross-correlations with and without
the mask of cluster regions.
The difference between two signals corresponds to the contribution from
massive clusters which are detected by \textit{Planck}.
Accordingly, the massive clusters can contribute to the signal
by at most $\simeq 20\%$, and thus
the rest of signal comes from the unresolved, i.e., low-mass, halos.
Hence, the cross-correlations contain the information from low-mass halos,
which are not easily accessible from other observables.

\begin{table*}
  \caption{The catalog of SZ detected clusters by \textit{Planck}
  which are located in HSC S16A footprints \citep{Medezinski2018}.
  The mass of clusters is inferred by fitting WL signal assuming the NFW profile.
  The positions of clusters are defined as those of brightest central galaxies.}
  \label{tab:cl_cat}
\begin{center}
\begin{tabular}{cccc}
Name in \textit{Planck} SZ catalog & NED name & RA (J2000) & Dec (J2000) \\
\hline \hline
PSZ2 G068.61-46.60 & Abell 2457 &
$22^\mathrm{h}35^\mathrm{m}40\fs80$ & $+01\degr29\arcmin05\farcs60$ \\
PSZ2 G167.98-59.95 & Abell 0329 &
$02^\mathrm{h}14^\mathrm{m}41\fs09$ & $-04\degr34\arcmin02\farcs46$ \\
PSZ2 G174.40-57.33 & Abell 0362 &
$02^\mathrm{h}31^\mathrm{m}41\fs17$ & $-04\degr52\arcmin57\farcs29$ \\
PSZ2 G228.50+34.95 & MaxBCG J140.53188+03.76632 &
$09^\mathrm{h}22^\mathrm{m}10\fs96$ & $+03\degr46\arcmin41\farcs52$ \\
PSZ2 G231.79+31.48 & MACS J0916.1-0023/Abell 0776 &
$09^\mathrm{h}16^\mathrm{m}09\fs24$ & $-00\degr24\arcmin16\farcs31$ \\
\hline
\end{tabular}
\begin{tabular}{cccc}
Redshift $z$ & Mass $M_\mathrm{200}$ [$10^{14} \, h^{-1} \, \Msun$] &
Radius $R_\mathrm{200}$ [$h^{-1} \, \mathrm{Mpc}$] & Angular extent $\theta_{200}$ [$\mathrm{arcmin}$] \\
\hline \hline
$0.0594$ & $2.02$ & $0.938$ & $19.43$ \\
$0.1393$ & $2.21$ & $0.943$ & $9.12$ \\
$0.1843$ & $4.12$ & $1.144$ & $8.79$ \\
$0.2701$ & $31.03$ & $2.177$ & $12.48$ \\
$0.3324$ & $8.10$ & $1.361$ & $6.75$ \\
\hline
\end{tabular}
\end{center}
\end{table*}

\subsection{Comparison with mass calibration measurements}
\label{sec:comp_obs}
Here, we discuss the implications of the results, especially the constraints on
the hydrostatic bias.
The constraints on the hydrostatic bias parameter is summarized
in Table~\ref{tab:b_constraints}.
As we have seen, the constraints on the bias parameter strongly depend on priors.
Here, we focus on the result with the \textit{Planck} prior because
previous works of mass calibration measurements and hydrodynamical simulations
also adopt \textit{Planck} cosmology or similar one.

\begin{figure*}
\includegraphics[width=\textwidth]{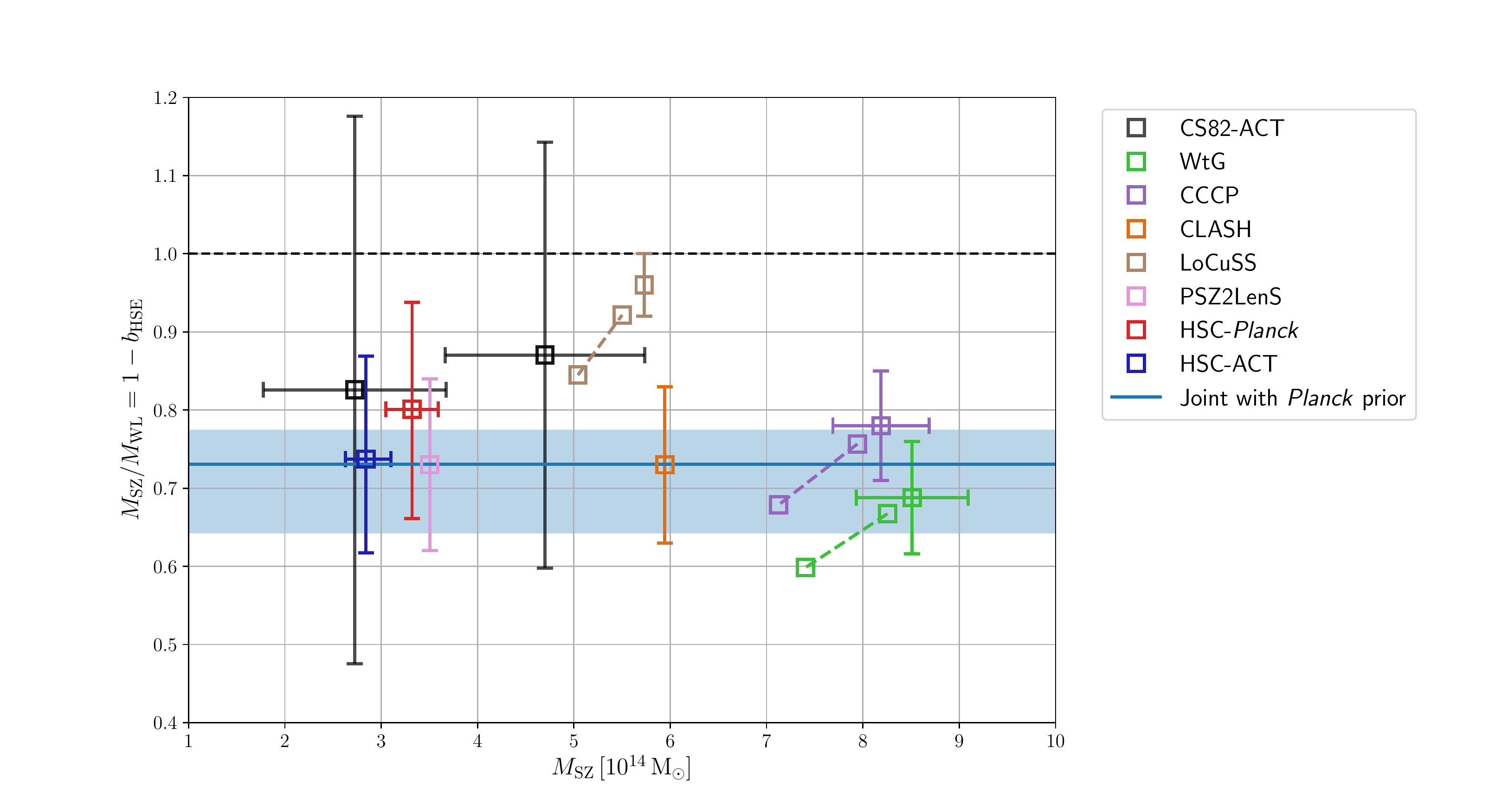}
\caption{The comparison of hydrostatic bias parameters
with respect to the hydrostatic mass $M_\mathrm{SZ}$
obtained with WL mass calibration measurements.
The results with the joint data set of the tSZ auto-power spectrum
and the tSZ-WL cross-correlations with the \textit{Planck} prior
is shown as blue bands, which correspond to 68\% confidence level,
and the blue solid line corresponds to the best-fit value.
The results in CS82-ACT \citep{Battaglia2016}, LoCuSS \citep{Smith2016},
CLASH \citep{Penna-Lima2017}, PSZ2LenS \citep{Sereno2017},
HSC-\textit{Planck} \citep{Medezinski2018} and
HSC-ACT \citep{Miyatake2019} are shown
as black, brown, orange, pink, red, and blue squares, respectively.
The green and purple squares show the results in WtG \citep{vonderLinden2014}
and CCCP \citep{Hoekstra2015}, respectively, and the same colored squares
connected with dashed lines are results where the Eddington bias are corrected
by the level of $3\text{--}15\%$ computed in \citet{Battaglia2016}.
In the LoCuSS measurement, the Eddington bias is estimated in \citet{vonderLinden2014}
and the corrected result is shown as the brown square connected with dashed lines.}
\label{fig:b_summary}
\end{figure*}

When the auto only or joint data sets are used,
the resultant constraint on the hydrostatic bias parameter is $b_\mathrm{HSE} \simeq 0.3$,
which is consistent with the hydrostatic mass bias of $0.1\text{--}0.3$
derived using X-ray/SZ and WL mass measurements of
individual clusters.\footnote{There might be covariance between
the secondary halo properties, e.g., halo shape or dynamical state,
and the signals from the galaxy cluster sample selected
by WL or tSZ \citep[see, e.g.,][]{Shirasaki2016}.
That might lead to bias in mass estimation by $10\text{--}20\%$.}
In the case with the cross only data set,
a slightly higher hydrostatic bias parameter ($b_\mathrm{HSE} \simeq 0.32$) is preferred,
although the error is large.
We note that the analysis of the tSZ auto-power spectrum by \citet{Bolliet2018}
also suggests a higher hydrostatic bias of $\simeq 40\%$.
We compare the obtained results and the previous mass calibration measurements
in Figure~\ref{fig:b_summary}.
Note that the tSZ auto-power spectrum and tSZ-WL cross-correlation are sensitive to
wide ranges of the halo mass (see Section~\ref{sec:differential_signal})
and thus the hydrostatic mass bias is not uniformly constrained with respect to mass.

One of the possibilities to explain the discrepancy between cross only and other data sets
is varying sensitivity to the halo mass and redshift range of these measurements.
Although the signal of tSZ auto-power spectrum comes
from galaxy clusters and groups with a wide range of mass \citep[see, e.g.,][]{Makiya2018},
mass calibration measurements probe only into massive clusters.
As Figure~\ref{fig:dCl} shows,
the tSZ-WL cross-correlation is sensitive to the structures at higher redshifts
compared to the tSZ auto-power spectrum \citep[see also][]{Battaglia2015}.
Thus, the discrepancy demonstrates non-thermal pressure depends on redshift or mass.
This hypothesis will be confirmed once the cross-correlations are
measured for larger areas and the constraints become tighter.

\subsection{Comparison with simulation predictions}
\label{sec:comp_sim}

Current observational constraints on the hydrostatic mass bias are
broadly consistent with the predictions of
hydrodynamical simulations \citep[e.g.,][]{Nagai2007b,Lau2009,Nelson2014a,Shi2016,Biffi2016,Henson2017}.
However, we emphasize that the predicted hydrostatic mass bias in the literature ranges
from 15 to 40\% depending the halo mass, and even depending on the
numerical codes and methods used.
Further studies of the nature and origin of the hydrostatic mass bias are clearly required.
In particular, it is important to understand the non-thermal pressure
contributing to the HSE equation \citep{Lau2013},
the effects of mass- and code-dependent temperature inhomogeneities effect
on the X-ray spectral temperature \citep{Rasia2014},
the role of acceleration term \citep{Suto2013,Nelson2014a},
and the redshift evolution of the non-thermal pressure and the HSE mass bias.

\begin{table}
\caption{Constraints on the hydrostatic parameter for each data set.
The central values are best-fit in each analysis and the errors correspond to $68\%$ C.L.,
which are estimated from parameter chains.}
\label{tab:b_constraints}
\begin{center}
\begin{tabular}{c|cc}
Data set & LSS prior & \textit{Planck} prior \\
\hline
\hline
Auto only & $0.306^{+0.001}_{-2.268}$ & $0.237^{+0.118}_{-0.010}$ \\
Cross only & $0.334^{+0.433}_{-0.390}$ & $0.320^{+0.437}_{-0.236}$ \\
Joint & $0.260^{+0.380}_{-0.105}$ & $0.269^{+0.089}_{-0.044}$ \\
\hline
\end{tabular}
\end{center}
\end{table}

\section{Conclusions}
\label{sec:conclusions}
We present measurements of cross-correlations of WL and tSZ
from HSC S16A and \textit{Planck} data
and derive constraints on cosmological parameters and hydrostatic bias parameter
using the tSZ auto-power spectrum and the tSZ-WL cross-correlations.
For WL, we reconstruct the convergence field
from HSC S16A shape catalog \citep{Mandelbaum2018a},
which covers $136.9 \, \mathrm{deg}^2$ with
the mean number density $n_g = 24.6 \, \mathrm{arcmin}^{-2}$.
For tSZ, we make use of the Compton-$y$ map based on \texttt{MILCA} algorithm
from $30$ to $857$ GHz channel maps of \textit{Planck} data \citep{Planck2015XXII}.
To calculate the auto-power spectrum and cross-correlation,
we use the halo model prescription with the universal pressure profile \citep{Nagai2007b}.
To calibrate the universal pressure profile,
the clusters are assumed to be in HSE, and thus the true mass may be
larger than the estimated mass under HSE because non-thermal pressure
support by turbulent motions and
other physical processes may be strong.
In order to account for the non-thermal pressure support,
we introduce the hydrostatic mass bias parameter $b_\mathrm{HSE}$,
which denotes the fraction of mass supported by non-thermal pressure,
and rescale the mass and the radius in the universal pressure profile.
For accurate estimation of covariance matrix,
We create realistic mock tSZ maps from all-sky $N$-body simulations \citep{Takahashi2017}
and use them to estimate the data covariance matrix accurately.
In addition to the mock tSZ maps, we also utilize the mock shape catalog,
which is created from the HSC S16A shape catalog \citep{Shirasaki2019b}.
The various systematic effects specific to the HSC observation, e.g.,
survey masks and discrete distribution of source galaxies, are incorporated in a direct manner.
Then, we compute the auto-power spectrum and cross-correlation from the suite of
mock maps, and estimate the covariance matrix.
Using the observational data and the covariance matrix estimated
from our 2268 mock catalogues,
we perform statistical analysis to constrain the cosmological parameters and
the hydrostatic bias parameter with the tSZ auto-power spectrum
and the tSZ-WL cross-correlation.
We add priors on cosmological parameters from
the combinations of results from measurements of JLA \citep{Betoule2014},
BOSS \citep{Alam2017}, and HSC cosmic shear analysis \citep{Hikage2019}, or
\textit{Planck} 2018 results of the temperature and polarization anisotropies of CMB
and CMB lensing \citep{Planck2018VI,Planck2018VIII}.
The hydrostatic bias parameter is strongly degenerate with cosmological parameters,
and thus the constraints depend on the choice of the priors.
In the case of using data sets with tSZ auto-power spectrum only or
joint analysis of tSZ auto-power spectrum and tSZ-WL cross-correlations
with the \textit{Planck} prior,
we find a reasonable value of the hydrostatic bias parameter $\simeq 30\%$,
which is consistent with WL mass calibration measurements \citep[e.g.,][]{Miyatake2019}
and the joint analysis of power spectra of cosmic shear with HSC and
tSZ with \textit{Planck} \citep{Makiya2019}.
On the other hand, when the data set only with tSZ-WL cross-correlations is employed,
slightly higher hydrostatic bias parameter ($\simeq 32\%$) is estimated.
Because both of tSZ power spectrum and tSZ-WL cross-correlations can probe into
less massive halos ($\lesssim 10^{14} \, \Msun$),
which are not accessible both for mass calibration measurements and hydrodynamical simulations,
the higher value of the hydrostatic bias can be realized by
high non-thermal pressure support in such less massive halos.
Since the HSC regions are limited to small sky coverage of $\sim 100 \, \mathrm{deg}^2$,
the constraint is not so tight, but for full sky coverage $\sim 1000\, \mathrm{deg}^2$,
tighter constraints can be obtained, and it will be possible to probe into
the redshift evolution of non-thermal pressure support via inference of hydrostatic bias
by a tomographic technique.
Furthermore, the ground-based CMB observatories, e.g., ACT and SPT,
are operating and Stage-IV CMB experiments will start operation in the near future.
The tSZ observations with high resolution and image quality by these experiments enable one
to utilize small scale ($< 1 \, \mathrm{arcmin}$) measurements,
which are useful to address the fine structure of the pressure profile of galaxy clusters.
Considering the large overlapping areas with HSC,
it is expected that we can fully trace the redshift evolution and mass dependence
of the non-thermal pressure at high precision with the upcoming ground-based CMB experiments.

\section*{Acknowledgements}
We acknowledge Nick Battaglia, Takashi Hamana, Chiaki Hikage, Eiichiro Komatsu, Takahiro Nishimichi,
Surhud More, Yasushi Suto, and Masahiro Takada for useful discussions.
KO is supported by Advanced Leading Graduate Course for Photon Science,
Research Fellowships of the Japan Society for the Promotion of Science (JSPS) for Young Scientists,
and JSPS Overseas Research Fellowships.
This work is supported by JSPS Grant-in-Aid for JSPS Research Fellow Grant Number JP16J01512 (KO),
JSPS KAKENHI Grant Numbers JP15H05893, JP17H01131 (RT),
JP15H05892, JP18K03693 (MO),
and by JST CREST Grant Number JPMJCR1414.
Numerical simulations were carried out on Cray XC50 at the Center for Computational Astrophysics,
National Astronomical Observatory of Japan.

The Hyper Suprime-Cam (HSC) collaboration includes
the astronomical communities of Japan and Taiwan, and Princeton University.
The HSC instrumentation and software were developed by the National Astronomical Observatory of Japan (NAOJ),
the Kavli Institute for the Physics and Mathematics of the Universe (Kavli IPMU),
the University of Tokyo, the High Energy Accelerator Research Organization (KEK),
the Academia Sinica Institute for Astronomy and Astrophysics in Taiwan (ASIAA), and Princeton University.
Funding was contributed by the FIRST program from Japanese Cabinet Office,
the Ministry of Education, Culture, Sports, Science and Technology (MEXT),
the Japan Society for the Promotion of Science (JSPS),
Japan Science and Technology Agency (JST), the Toray Science Foundation,
NAOJ, Kavli IPMU, KEK, ASIAA, and Princeton University.

This paper makes use of software developed for the Large Synoptic Survey Telescope.
We thank the LSST Project for making their code available as free software at \url{http://dm.lsst.org}.

Based on data collected at the Subaru Telescope and retrieved from the HSC data archive system,
which is operated by the Subaru Telescope and Astronomy Data Center
at National Astronomical Observatory of Japan.



\bibliographystyle{mnras}
\bibliography{main}



\appendix

\section{Effects of choice of algorithms in photometric redshifts estimation}
\label{sec:pz_algos}
\begin{figure}
\includegraphics[width=\columnwidth]{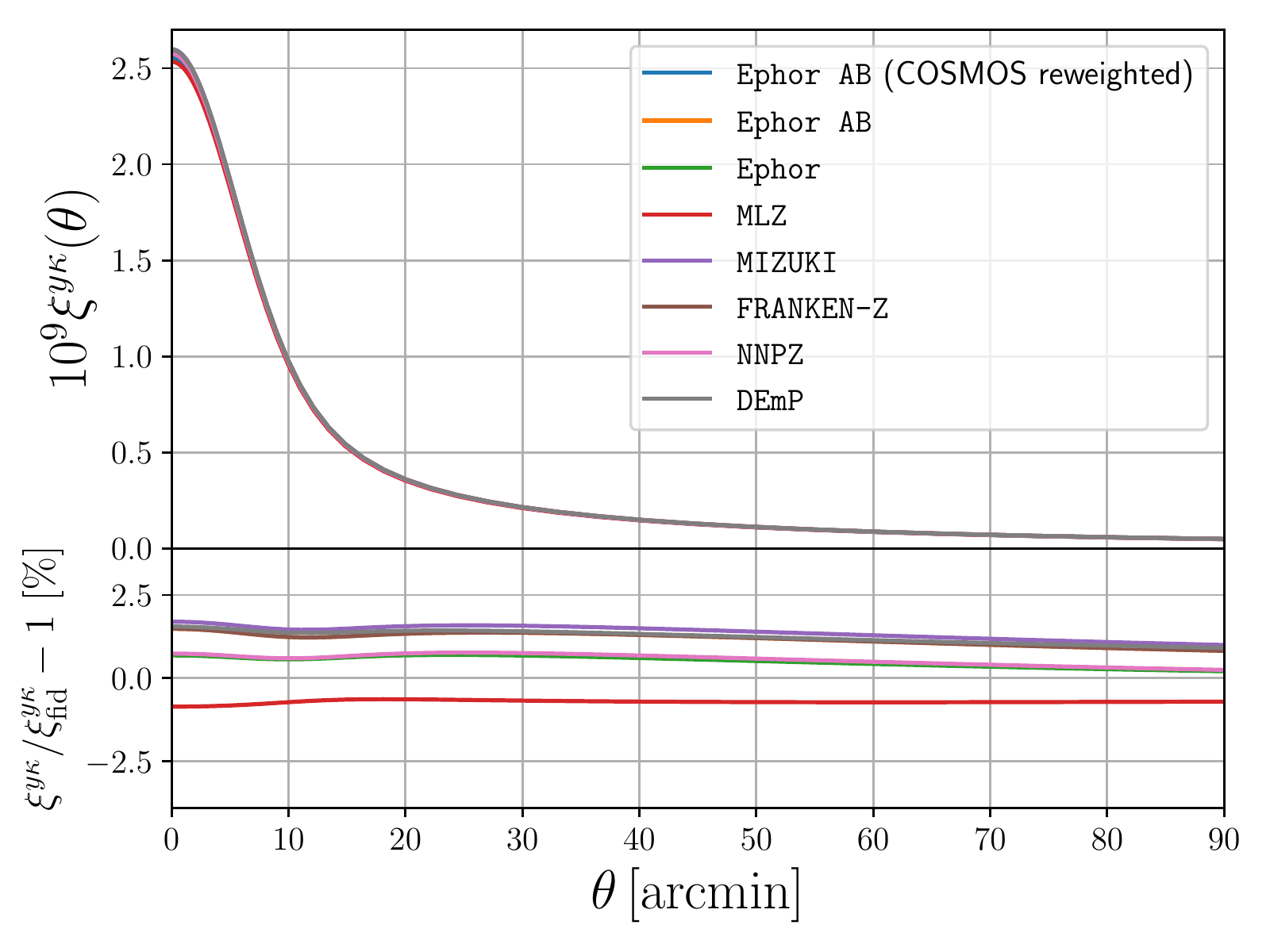}
\caption{The cross-correlations based on halo model
with different codes of photometric redshift estimation.
The lower panel shows the fractional difference from
the fiducial cross-correlation, i.e., the one with \texttt{Ephor AB}
with reweights of COSMOS 30-band observations.}
\label{fig:pz_yk}
\end{figure}

We show the dependence on algorithms of photometric redshift estimation
for the cross-correlation calculations.
Figure~\ref{fig:pz_yk} shows the halo model calculations
with \textit{WMAP} 9-yr cosmological parameters (see Section~\ref{sec:mock_observations})
for eight different stacked PDFs.
The difference from the fiducial model, i.e., the one with \texttt{Ephor AB}
with reweights of COSMOS 30-band observations, is within $2\%$ at all scales.

\section{Estimator of inverse covariance matrix}
\label{sec:invcov}
In the case of dealing with correlated and high dimensional data,
to compute the inverse covariance matrix,
standard inversion of estimated covariance matrix $S$ may lead to
numerically unstable estimation of the inverse covariance matrix.
For the sparse matrix, the graphical LASSO algorithm can
efficiently estimate the inverse covariance matrix $\Sigma$ with $L_1$ regularization:
\beq
\hat{\Sigma} = \argmax_\Sigma \left( \log \mathrm{det} \Sigma - \mathrm{Tr} (S \Sigma)
-\lambda \sum_{i \neq j} |\Sigma_{ij}| \right) ,
\eeq
where the last term is the penalty term and $\lambda$ is the regularizaion parameter.
We employ the \texttt{GraphicalLassoCV} package in \texttt{scikit-learn} \citep{scikit-learn}
and determine the parameter $\lambda$ with 5-fold cross-validation method.

There is a caveat for estimation of the inverse covariance matrix
for the joint analysis of the tSZ auto-power spectrum and the tSZ-WL cross-correlation.
If graphical LASSO is applied to the full covariance matrix to obtain
the full inverse covariance matrix, it may lead to excessive suppression of off-diagonal terms.
Thus, we compute the inverse covariance matrix with the block-wise inversion
and keep the leading terms with respect to the cross-covariance
because the cross-covariance is much smaller than the auto-covariance.
The resultant inverse covariance matrix is given as
\beq
\begin{pmatrix}
S_{C C} & S_{C \xi} \\
S_{\xi C} & S_{\xi \xi}
\end{pmatrix}^{-1}
\simeq
\begin{pmatrix}
\Sigma_{CC} & -\Sigma_{CC} S_{C\xi} \Sigma_{\xi\xi} \\
-\Sigma_{\xi\xi} S_{\xi C} \Sigma_{CC} & \Sigma_{\xi\xi}
\end{pmatrix}
,
\eeq
where $S$ is the covariance matrix, the subscript denotes the employed data,
and $S_{\xi C} = S_{C \xi}^T$.
In order to obtain the full inverse covariance matrix,
we need to compute the inverse matrices, $\Sigma_{CC}$ and $\Sigma_{\xi\xi}$,
with graphical LASSO instead of inversion of the full covariance matrix.

\section{Constraints for full parameter space}
\label{sec:full_space}
\begin{figure*}
\includegraphics[width=\textwidth]{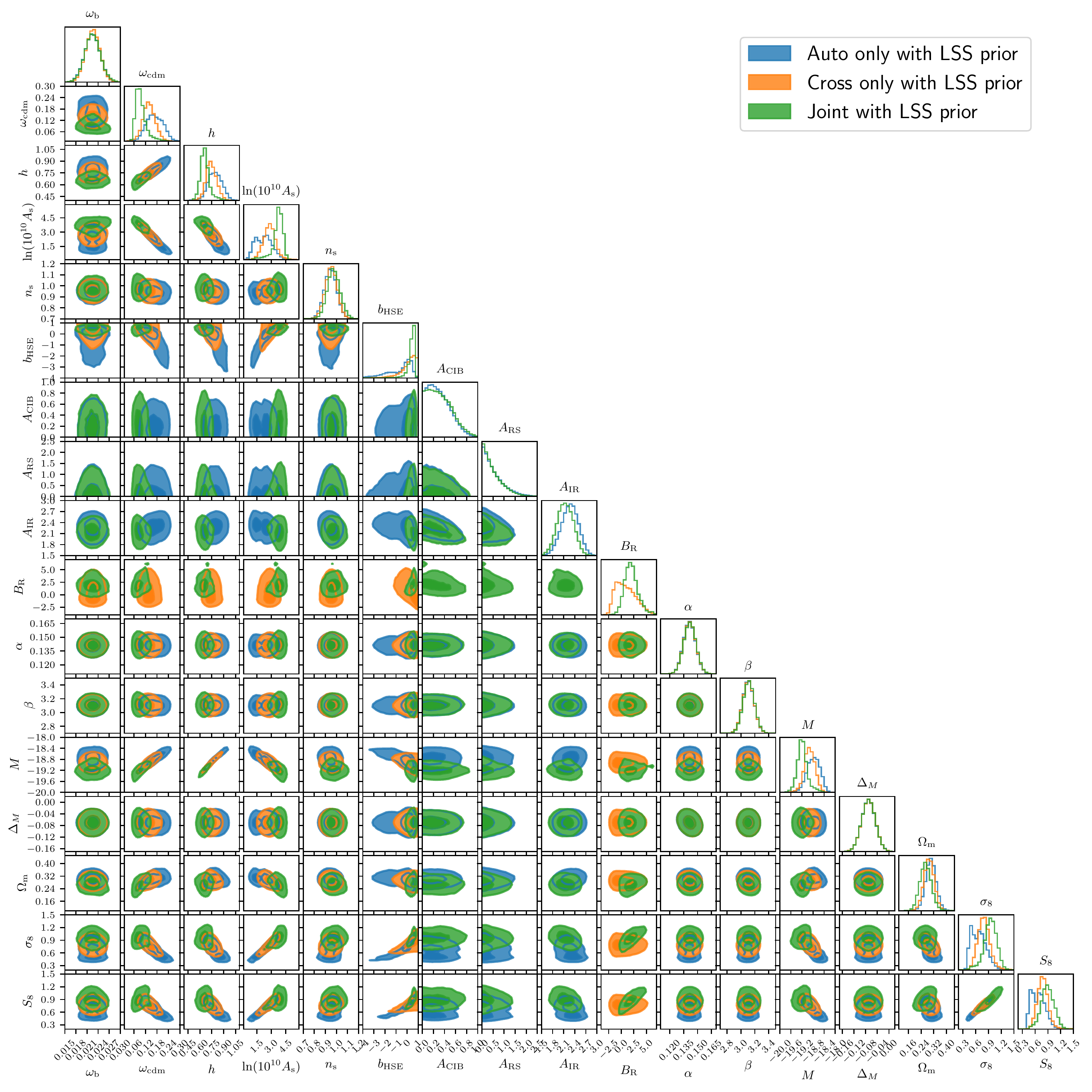}
\caption{Confidence regions of cosmological parameters, hydrostatic bias parameter
and all nuisance parameters inferred with three data sets (auto only, cross only, and joint)
with the LSS prior. The inner (outer) contour corresponds to the $1\sigma$ ($2\sigma$) level.
The diagonal panels show the marginalized likelihoods.}
\label{fig:fulltriangle_LSS}
\end{figure*}

\begin{figure*}
\includegraphics[width=\textwidth]{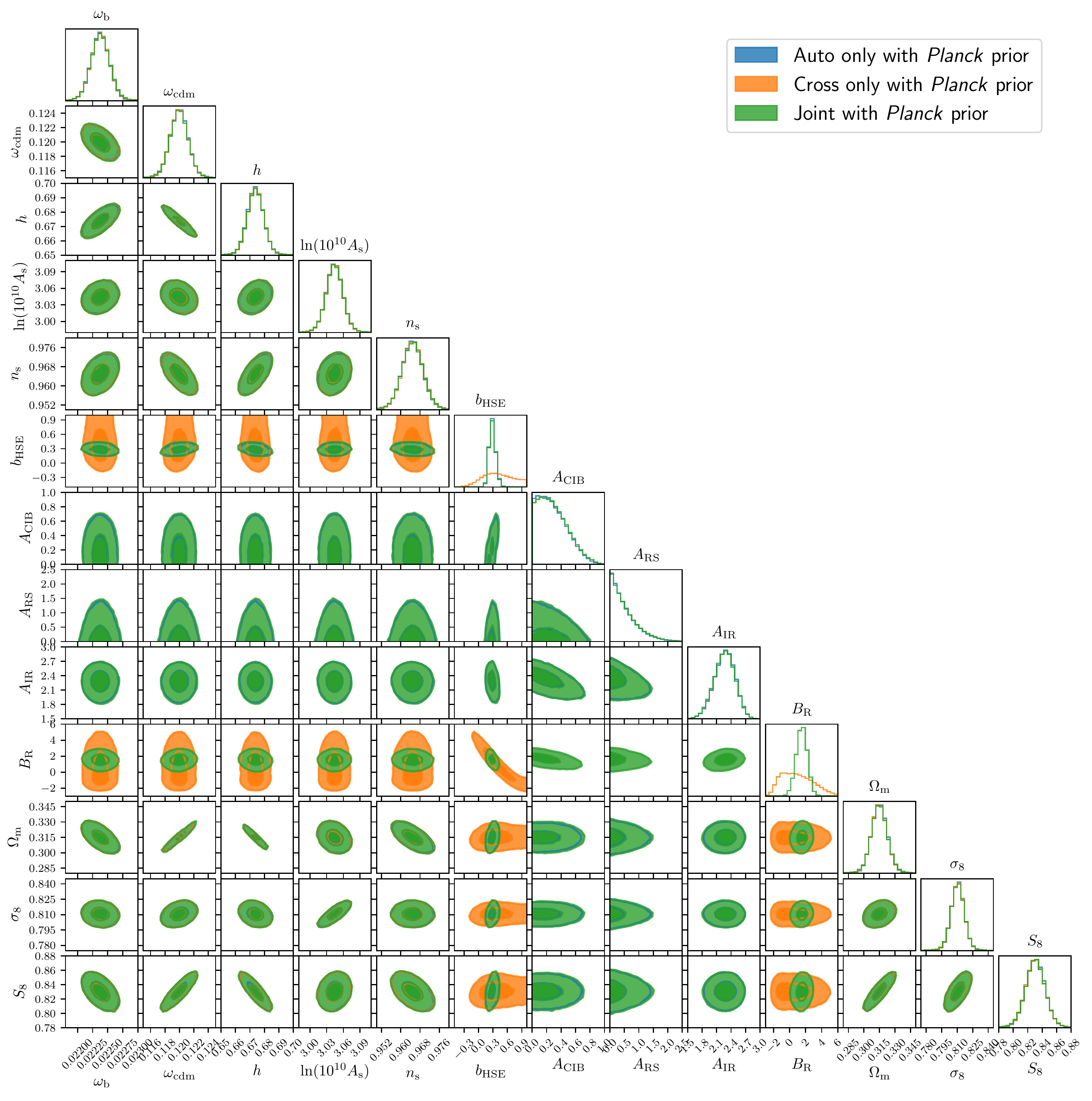}
\caption{Confidence regions of cosmological parameters and hydrostatic bias parameter
and all nuisance parameters inferred with three data sets (auto only, cross only, and joint)
with the \textit{Planck} prior. The inner (outer) contour corresponds to the $1\sigma$ ($2\sigma$) level.
The diagonal panels show the marginalized likelihoods.}
\label{fig:fulltriangle_Plancklensing}
\end{figure*}

Here, we show the confidence regions on all parameters in Figures~\ref{fig:fulltriangle_LSS}
and \ref{fig:fulltriangle_Plancklensing} for LSS prior and \textit{Planck} prior, respectively.
In Tables~\ref{tab:constraints_LSS} and \ref{tab:constraints_Plancklensing},
best-fit values and marginalized errors of all parameters
for LSS prior and \textit{Planck} prior, respectively, are shown.
For the LSS prior, in addition to the nuisance parameters which determine amplitudes of
foreground contributions in the tSZ auto-power spectrum and the tSZ-WL cross-correlations,
there are four more nuisance parameters, $\alpha$, $\beta$, $M$, and $\Delta_M$,
in the magnitude-magnification relation of the JLA analysis,
which definitions are found in Eqs.~(4) and (5) of \citet{Betoule2014}.

\begin{table*}
\caption{Best-fit and median values estimated from parameter chains
and marginalized constraints ($68\%$ C.L. and $95\%$ C.L.) on all parameters inferred from
three data sets (auto only, cross only, and joint) with the LSS prior.
The last three parameters ($\Omega_\mathrm{m}$, $\sigma_8$, and $S_8$) are derived parameters.}
\label{tab:constraints_LSS}
\begin{center}
\begin{tabular}{cccccc}
\multicolumn{6}{c}{\textbf{Auto only}} \\
Parameter & Range & Best-fit & Median & $68\%$ C.L. & $95\%$ C.L. \\
\hline \hline
$\omega_\mathrm{b}$ & --- & $0.02226$ & $0.02244$ & $[0.02036,0.02451]$ & $[0.01822,0.02653]$ \\
$\omega_\mathrm{cdm}$ & --- & $0.1215$ & $0.1664$ & $[0.1232,0.2133]$ & $[0.08282,0.2488]$ \\
$h$ & --- & $0.6932$ & $0.7906$ & $[0.7072,0.8790]$ & $[0.6239,0.9548]$ \\
$\ln (10^{10} A_\mathrm{s})$ & --- & $3.140$ & $2.244$ & $[1.292,3.093]$ & $[0.7963,3.889]$ \\
$n_\mathrm{s}$ & --- & $0.9304$ & $0.9470$ & $[0.8914,1.004]$ & $[0.8344,1.063]$ \\
$b_\mathrm{HSE}$ & $(-\infty, 1]$ & $0.3057$ & $-0.2889$ & $[-1.951,0.3169]$ & $[-3.514,0.5868]$ \\
$A_\mathrm{CIB}$ & $[0, 10]$ & $0.1816$ & $0.2768$ & $[0.09240,0.5177]$ & $[0.01450,0.7524]$ \\
$A_\mathrm{RS}$ & $[0, 10]$ & $0.2313$ & $0.4124$ & $[0.1087,0.9969]$ & $[0.01561,1.780]$ \\
$A_\mathrm{IR}$ & $[0, 10]$ & $2.414$ & $2.274$ & $[2.035,2.496]$ & $[1.800,2.696]$ \\
$\alpha$ & --- & $0.1436$ & $0.1412$ & $[0.1346,0.1478]$ & $[0.1280,0.1545]$ \\
$\beta$ & --- & $3.085$ & $3.105$ & $[3.025,3.187]$ & $[2.946,3.270]$ \\
$M$ & --- & $-19.07$ & $-18.78$ & $[-19.02,-18.55]$ & $[-19.31,-18.38]$ \\
$\Delta_M$ & --- & $-0.07714$ & $-0.07014$ & $[-0.09346,-0.04660]$ & $[-0.1171,-0.02307]$ \\
\hline
$\Omega_\mathrm{m}$ & --- & $0.3005$ & $0.3010$ & $[0.2684,0.3339]$ & $[0.2320,0.3683]$ \\
$\sigma_8$ & --- & $0.8528$ & $0.6687$ & $[0.4930,0.8576]$ & $[0.4228,1.056]$ \\
$S_8$ & --- & $0.8536$ & $0.6698$ & $[0.5030,0.8395]$ & $[0.4337,1.006]$ \\
\hline
\end{tabular}

\begin{tabular}{cccccc}
\multicolumn{6}{c}{\textbf{Cross only}} \\
Parameter & Range & Best-fit & Median & $68\%$ C.L. & $95\%$ C.L. \\
\hline \hline
$\omega_\mathrm{b}$ & --- & $0.02255$ & $0.02256$ & $[0.02065,0.02447]$ & $[0.01872,0.02642]$ \\
$\omega_\mathrm{cdm}$ & --- & $0.1177$ & $0.1380$ & $[0.1077,0.1707]$ & $[0.07441,0.2059]$ \\
$h$ & --- & $0.6932$ & $0.7411$ & $[0.6821,0.8124]$ & $[0.6139,0.8871]$ \\
$\ln (10^{10} A_\mathrm{s})$ & --- & $3.201$ & $2.816$ & $[2.187,3.419]$ & $[1.547,4.242]$ \\
$n_\mathrm{s}$ & --- & $0.9444$ & $0.9537$ & $[0.9002,1.007]$ & $[0.8445,1.063]$ \\
$b_\mathrm{HSE}$ & $(-\infty, 1]$ & $0.3339$ & $0.3012$ & $[-0.5595,0.7667]$ & $[-2.261,0.9651]$ \\
$B_\mathrm{R}$ & $[-10, 10]$ & $1.679$ & $0.7335$ & $[-1.077,3.090]$ & $[-2.020,5.333]$ \\
$\alpha$ & --- & $0.1393$ & $0.1414$ & $[0.1348,0.1481]$ & $[0.1282,0.1549]$ \\
$\beta$ & --- & $3.087$ & $3.107$ & $[3.026,3.188]$ & $[2.946,3.272]$ \\
$M$ & --- & $-19.08$ & $-18.92$ & $[-19.11,-18.73]$ & $[-19.35,-18.54]$ \\
$\Delta_M$ & --- & $-0.06513$ & $-0.06980$ & $[-0.09318,-0.04654]$ & $[-0.1167,-0.02346]$ \\
\hline
$\Omega_\mathrm{m}$ & --- & $0.2932$ & $0.2901$ & $[0.2578,0.3225]$ & $[0.2239,0.3559]$ \\
$\sigma_8$ & --- & $0.8636$ & $0.7929$ & $[0.6584,0.9376]$ & $[0.5352,1.156]$ \\
$S_8$ & --- & $0.8538$ & $0.7775$ & $[0.6569,0.9015]$ & $[0.5453,1.089]$ \\
\hline
\end{tabular}

\begin{tabular}{cccccc}
\multicolumn{6}{c}{\textbf{Joint}} \\
Parameter & Range & Best-fit & Median & $68\%$ C.L. & $95\%$ C.L. \\
\hline \hline
$\omega_\mathrm{b}$ & --- & $0.02232$ & $0.02247$ & $[0.02044,0.02459]$ & $[0.01834,0.02689]$ \\
$\omega_\mathrm{cdm}$ & --- & $0.1225$ & $0.09042$ & $[0.07107,0.1208]$ & $[0.05830,0.1834]$ \\
$h$ & --- & $0.7080$ & $0.6580$ & $[0.6122,0.7121]$ & $[0.5590,0.8294]$ \\
$\ln (10^{10} A_\mathrm{s})$ & --- & $3.098$ & $3.778$ & $[3.297,4.169]$ & $[1.827,4.468]$ \\
$n_\mathrm{s}$ & --- & $0.9365$ & $0.9740$ & $[0.9192,1.033]$ & $[0.8619,1.095]$ \\
$b_\mathrm{HSE}$ & $(-\infty, 1]$ & $0.2600$ & $0.4755$ & $[0.1548,0.6401]$ & $[-0.9675,0.7336]$ \\
$A_\mathrm{CIB}$ & $[0, 10]$ & $0.1749$ & $0.2973$ & $[0.09437,0.5483]$ & $[0.01332,0.7999]$ \\
$A_\mathrm{RS}$ & $[0, 10]$ & $0.3367$ & $0.3847$ & $[0.1019,0.9350]$ & $[0.01393,1.623]$ \\
$A_\mathrm{IR}$ & $[0, 10]$ & $2.404$ & $2.143$ & $[1.926,2.367]$ & $[1.707,2.568]$ \\
$B_\mathrm{R}$ & $[-10, 10]$ & $2.088$ & $1.913$ & $[0.7250,3.299]$ & $[-0.3605,5.322]$ \\
$\alpha$ & --- & $0.1433$ & $0.1416$ & $[0.1350,0.1484]$ & $[0.1283,0.1551]$ \\
$\beta$ & --- & $3.094$ & $3.113$ & $[3.031,3.194]$ & $[2.951,3.277]$ \\
$M$ & --- & $-19.02$ & $-19.20$ & $[-19.35,-19.02]$ & $[-19.54,-18.68]$ \\
$\Delta_M$ & --- & $-0.07639$ & $-0.06996$ & $[-0.09347,-0.04679]$ & $[-0.1165,-0.02393]$ \\
\hline
$\Omega_\mathrm{m}$ & --- & $0.2903$ & $0.2680$ & $[0.2327,0.3047]$ & $[0.1996,0.3453]$ \\
$\sigma_8$ & --- & $0.8445$ & $0.9531$ & $[0.8083,1.090]$ & $[0.5673,1.218]$ \\
$S_8$ & --- & $0.8307$ & $0.8904$ & $[0.7462,1.045]$ & $[0.5680,1.185]$ \\
\hline
\end{tabular}
\end{center}
\end{table*}

\begin{table*}
\caption{Best-fit and median values estimated from parameter chains
and marginalized constraints ($68\%$ C.L. and $95\%$ C.L.) on all parameters inferred from
three data sets (auto only, cross only, and joint) with the \textit{Planck} prior.
The last three parameters ($\Omega_\mathrm{m}$, $\sigma_8$, and $S_8$) are derived parameters.}
\label{tab:constraints_Plancklensing}
\begin{center}
\begin{tabular}{cccccc}
\multicolumn{6}{c}{\textbf{Auto only}} \\
Parameter & Range & Best-fit & Median & $68\%$ C.L. & $95\%$ C.L. \\
\hline \hline
$\omega_\mathrm{b}$ & --- & $0.02233$ & $0.02237$ & $[0.02222,0.02253]$ & $[0.02206,0.02268]$ \\
$\omega_\mathrm{cdm}$ & --- & $0.1201$ & $0.1200$ & $[0.1187,0.1212]$ & $[0.1175,0.1224]$ \\
$h$ & --- & $0.6729$ & $0.6737$ & $[0.6682,0.6793]$ & $[0.6624,0.685]$ \\
$\ln (10^{10} A_\mathrm{s})$ & --- & $3.044$ & $3.044$ & $[3.029,3.059]$ & $[3.013,3.075]$ \\
$n_\mathrm{s}$ & --- & $0.9655$ & $0.9649$ & $[0.9606,0.9691]$ & $[0.9560,0.9737]$ \\
$b_\mathrm{HSE}$ & $(-\infty, 1]$ & $0.2367$ & $0.2849$ & $[0.2266,0.3548]$ & $[0.1745,0.4241]$ \\
$A_\mathrm{CIB}$ & $[0, 10]$ & $0.1171$ & $0.2601$ & $[0.08309,0.4952]$ & $[0.01245,0.7257]$ \\
$A_\mathrm{RS}$ & $[0, 10]$ & $0.06770$ & $0.3941$ & $[0.1039,0.9455]$ & $[0.01391,1.717]$ \\
$A_\mathrm{IR}$ & $[0, 10]$ & $2.522$ & $2.274$ & $[2.051,2.479]$ & $[1.820,2.663]$ \\
\hline
$\Omega_\mathrm{m}$ & --- & $0.3160$ & $0.3151$ & $[0.3075,0.3227]$ & $[0.3000,0.3308]$ \\
$\sigma_8$ & --- & $0.8115$ & $0.8106$ & $[0.8043,0.8168]$ & $[0.7976,0.8237]$ \\
$S_8$ & --- & $0.8329$ & $0.8307$ & $[0.8173,0.844]$ & $[0.8035,0.8582]$ \\
\hline
\end{tabular}

\begin{tabular}{cccccc}
\multicolumn{6}{c}{\textbf{Cross only}} \\
Parameter & Range & Best-fit & Median & $68\%$ C.L. & $95\%$ C.L. \\
\hline \hline
$\omega_\mathrm{b}$ & --- & $0.02240$ & $0.02238$ & $[0.02222,0.02253]$ & $[0.02204,0.02271]$ \\
$\omega_\mathrm{cdm}$ & --- & $0.1201$ & $0.1199$ & $[0.1187,0.1212]$ & $[0.1172,0.1227]$ \\
$h$ & --- & $0.6734$ & $0.6738$ & $[0.6682,0.6795]$ & $[0.6613,0.6863]$ \\
$\ln (10^{10} A_\mathrm{s})$ & --- & $3.045$ & $3.044$ & $[3.029,3.059]$ & $[3.010,3.076]$ \\
$n_\mathrm{s}$ & --- & $0.9648$ & $0.9650$ & $[0.9606,0.9694]$ & $[0.9554,0.9745]$ \\
$b_\mathrm{HSE}$ & $(-\infty, 1]$ & $0.3195$ & $0.3932$ & $[0.08367,0.7564]$ & $[-0.1732,0.9653]$ \\
$B_\mathrm{R}$ & $[-10, 10]$ & $1.289$ & $0.8204$ & $[-1.040,3.115]$ & $[-2.023,5.325]$ \\
\hline
$\Omega_\mathrm{m}$ & --- & $0.3157$ & $0.3148$ & $[0.3072,0.3227]$ & $[0.2986,0.3322]$ \\
$\sigma_8$ & --- & $0.8115$ & $0.8105$ & $[0.8041,0.8169]$ & $[0.7962,0.8245]$ \\
$S_8$ & --- & $0.8325$ & $0.8304$ & $[0.8168,0.8441]$ & $[0.8010,0.8601]$ \\
\hline
\end{tabular}

\begin{tabular}{cccccc}
\multicolumn{6}{c}{\textbf{Joint}} \\
Parameter & Range & Best-fit & Median & $68\%$ C.L. & $95\%$ C.L. \\
\hline \hline
$\omega_\mathrm{b}$ & --- & $0.02235$ & $0.02238$ & $[0.02222,0.02254]$ & $[0.02204,0.02274]$ \\
$\omega_\mathrm{cdm}$ & --- & $0.1201$ & $0.1199$ & $[0.1187,0.1212]$ & $[0.1170,0.1228]$ \\
$h$ & --- & $0.6728$ & $0.6738$ & $[0.6680,0.6796]$ & $[0.6609,0.6871]$ \\
$\ln (10^{10} A_\mathrm{s})$ & --- & $3.044$ & $3.044$ & $[3.029,3.059]$ & $[3.010,3.078]$ \\
$n_\mathrm{s}$ & --- & $0.9644$ & $0.9649$ & $[0.9604,0.9694]$ & $[0.9546,0.9746]$ \\
$b_\mathrm{HSE}$ & $(-\infty, 1]$ & $0.2692$ & $0.2857$ & $[0.2251,0.3578]$ & $[0.1695,0.4323]$ \\
$A_\mathrm{CIB}$ & $[0, 10]$ & $0.2876$ & $0.2691$ & $[0.08802,0.5089]$ & $[0.01381,0.7486]$ \\
$A_\mathrm{RS}$ & $[0, 10]$ & $0.04537$ & $0.4048$ & $[0.1057,0.9772]$ & $[0.01403,1.780]$ \\
$A_\mathrm{IR}$ & $[0, 10]$ & $2.425$ & $2.266$ & $[2.040,2.474]$ & $[1.812,2.663]$ \\
$B_\mathrm{R}$ & $[-10, 10]$ & $1.498$ & $1.522$ & $[0.8353,2.180]$ & $[0.1187,2.824]$ \\
\hline
$\Omega_\mathrm{m}$ & --- & $0.3161$ & $0.3149$ & $[0.3071,0.3228]$ & $[0.2976,0.3326]$ \\
$\sigma_8$ & --- & $0.8111$ & $0.8105$ & $[0.8040,0.8170]$ & $[0.7955,0.8250]$ \\
$S_8$ & --- & $0.8326$ & $0.8304$ & $[0.8166,0.8443]$ & $[0.7989,0.8612]$ \\
\hline
\end{tabular}
\end{center}
\end{table*}


\bsp 
\label{lastpage}
\end{document}